\documentclass[iop,apj,numberedappendix,twocolumn, tighten]{aastex62}

\usepackage{arydshln}
\usepackage{natbib}
\usepackage{graphicx}
\usepackage{amsmath,amsfonts}
\usepackage{multirow}
\usepackage{xspace}
\usepackage[normalem]{ulem}
\usepackage{booktabs}
\usepackage{colortbl}
\usepackage{hyperref}
\usepackage[nameinlink]{cleveref}
\usepackage[encapsulated]{CJK}
\usepackage{ucs}
\usepackage{float}
\usepackage[utf8x]{inputenc}

\usepackage{eht}
\usepackage{newtxtext}
\usepackage{tablefootnote} 
\usepackage{mathrsfs}
\usepackage{lineno}

\usepackage{array}
\definecolor{fed_blue}{HTML}{07004D}
\definecolor{steel_blue}{HTML}{2D82B7}
\definecolor{steel_blue_dark}{HTML}{1C71A6}
\definecolor{aqua_marine}{HTML}{42E2B8}
\definecolor{dutch_white}{HTML}{F3DFBF}
\definecolor{light_coral}{HTML}{EB8A90}
\definecolor{light_coral_dark}{HTML}{BA5A60}

\hypersetup{
    colorlinks = true,
    linkcolor = purple,
    urlcolor  = steel_blue_dark,
    citecolor = steel_blue_dark,
    anchorcolor = black
}

\newcommand{\hl}[1]{#1}
\newcommand{\hll}[1]{#1}

\graphicspath{{./figures}}

\def\Snospace~{\S{}}

\newcommand{\angstrom}{\AA\xspace}

\def\lsim{\mathrel{\raise.3ex\h box{$<$\kern-.75em\lower1ex\hbox{$\sim$}}}}
\def\gsim{\mathrel{\raise.3ex\hbox{$>$\kern-.75em\lower1ex\hbox{$\sim$}}}}
\def\gtwid{\mathrel{\raise.3ex\hbox{$>$\kern-.75em\lower1ex\hbox{$\sim$}}}}
\def\proptwid{\mathrel{\raise.3ex\hbox{$\propto$\kern-.75em\lower1ex\hbox{$\sim$}}}}

\defcitealias{TMS}{TMS}
\setcitestyle{notesep={; }}

\begin{document}

\title{\textbf{Shock-cooling Constraints via Early-time Observations of the Type IIb SN 2022hnt}} 
\shorttitle{SN 2022hnt Shock Cooling}

\author[0000-0003-4914-5625]{Joseph R. Farah}
\affiliation{Las Cumbres Observatory, 6740 Cortona Drive, Suite 102, Goleta, 
CA 93117-5575, USA}
\affiliation{Department of Physics, University of California, Santa Barbara, 
CA 93106-9530, USA}

\author[0000-0003-4253-656X]{D. Andrew Howell}
\affiliation{Las Cumbres Observatory, 6740 Cortona Drive, Suite 102, Goleta, 
CA 93117-5575, USA}
\affiliation{Department of Physics, University of California, Santa Barbara, 
CA 93106-9530, USA}

\author[0000-0003-0794-5982]{Giacomo Terreran}
\affiliation{Las Cumbres Observatory, 6740 Cortona Drive, Suite 102, Goleta, 
CA 93117-5575, USA}
\author[0000-0002-7996-8780]{Ido Irani}
\affiliation{Department of Particle Physics and Astrophysics,
             Weizmann Institute of Science,
             234 Herzl St, 7610001 Rehovot, Israel}
\author[0009-0002-8735-6298]{Jonathan Morag}
\affiliation{Department of Particle Physics and Astrophysics,
             Weizmann Institute of Science,
             234 Herzl St, 7610001 Rehovot, Israel}

\author[0000-0002-7472-1279]{Craig Pellegrino}
\affiliation{Department of Astronomy, University of Virginia, Charlottesville, VA 22904, USA}

\author[0000-0001-5807-7893]{Curtis McCully}
\affiliation{Las Cumbres Observatory, 6740 Cortona Drive, Suite 102, Goleta, 
CA 93117-5575, USA}

\author[0000-0001-9570-0584]{Megan Newsome}
\affiliation{Las Cumbres Observatory, 6740 Cortona Drive, Suite 102, Goleta, 
CA 93117-5575, USA}
\affiliation{Department of Physics, University of California, Santa Barbara, 
CA 93106-9530, USA}

\author[0000-0003-0209-9246]{Estefania Padilla Gonzalez}
\affiliation{Las Cumbres Observatory, 6740 Cortona Drive, Suite 102, Goleta, 
CA 93117-5575, USA}
\affiliation{Department of Physics, University of California, Santa Barbara, 
CA 93106-9530, USA}

\author[0000-0002-4924-444X]{Azalee Bostroem}
\affiliation{Steward Observatory, University of Arizona, 933 North Cherry Avenue, Tucson, AZ 85721-0065, USA}

\author[0000-0002-0832-2974]{Griffin Hosseinzadeh}
\affiliation{Steward Observatory, University of Arizona, 933 North Cherry Avenue, Tucson, AZ 85721-0065, USA}
\affiliation{Department of Astronomy \& Astrophysics, University of California, San Diego, 9500 Gilman Drive, MC 0424, La Jolla, CA 92093-0424, USA}

\author[0000-0002-1895-6639]{Moira Andrews}
\affiliation{Las Cumbres Observatory, 6740 Cortona Drive, Suite 102, Goleta, 
CA 93117-5575, USA}
\affiliation{Department of Physics, University of California, Santa Barbara, 
CA 93106-9530, USA}

\author[0000-0002-7174-8273]{Logan Prust}
\affiliation{Kavli Institute for Theoretical Physics, University of California, Santa Barbara, CA 93106, USA}
\author[0000-0002-1125-9187]{Daichi Hiramatsu}
\affiliation{Center for Astrophysics \textbar{} Harvard \& Smithsonian, 60 Garden Street, Cambridge, MA 02138-1516, USA} AND \affiliation{The NSF AI Institute for Artificial Intelligence and Fundamental Interactions, USA}

\shortauthors{Farah et al.}

\correspondingauthor{$^\dag$Joseph R. Farah}
\email{josephfarah@ucsb.edu}

\begin{abstract}
We report the results of a rapid follow-up campaign on the Type IIb Supernova (SN) 2022hnt. We present a daily, multi-band, photometric follow-up using the Las Cumbres Observatory, the Zwicky Transient Facility, the orbiting \textit{Swift} observatory, and the Asteroid Terrestrial-impact Last Alert System (ATLAS). 
A distinctive feature in the light curve of SN 2022hnt and other IIb SNe is an early narrow peak prior to the ${}^{56}$Ni peak caused by rapid shock cooling of the hydrogen envelope, which can serve as an important probe of the properties of the massive progenitor star in the moments before explosion. Using SN 2022hnt as a case study, we demonstrate a framework of considerations for the application of shock cooling models to type IIb SNe, outlining a consistent procedure for future surveys of Type IIb SNe progenitor and explosion properties. \hll{We fit several recent models of shock-cooling emission and obtain progenitor radii between $\sim50$ and $\sim100$ $R_\odot$, as well as hydrogen-enriched envelope masses between $\sim0.01$ and $\sim0.1$ $M_\odot$, both consistent with values for other IIb SNe. One of these models is the model of \cite{Morag2023}, marking the first time this model has been applied to a Type IIb SN.} We evaluate contrasting predictions between shock-cooling models to construct a fiducial parameter set which can be used for comparison to other SNe. Finally, we investigate the possibility of extended wind breakout or precursor emission captured in the earliest detections.
\end{abstract}


\section{Introduction}
\label{sec:introduction}

Type IIb supernovae (SNe IIb) represent a subclass of core-collapse supernovae characterized by the presence of hydrogen lines in their early spectra, which subsequently disappear \citep{Woosley1994,Filippenko1997}. They are believed to originate from massive stars that have shed a substantial portion of their outer envelopes \citep[e.g.,][and references therein]{Sravan2019}. 
This extensive mass loss prior to explosion results in the removal of most, but not all, of the \hl{hydrogen-rich material in the progenitor's outer shell}. The mass loss mechanism in Type IIb SNe progenitors is not well-understood, but is thought to occur via multiple channels as a result of stellar wind stripping, accretion onto a binary companion, or rotation effects \citep{Smith2014,
Ouchi2017}. Observational and theoretical studies suggest that SNe IIb may arise from a variety of progenitor scenarios, including single massive stars and binary systems \citep[e.g.,][]{Claeys2011,Benvenuto2013,Smartt2015,Dyk2017,Modjaz2019,Niu2024}.  \hl{The diversity of progenitor and mass-loss channels for Type IIb SNe make them a useful probe of the characteristics of near-death massive stars}. This population of stars plays an important role in the evolution of galaxies and the metal enrichment of the interstellar medium \citep{Andrews2017}. Ongoing studies continue to refine our understanding of the progenitor systems and explosion mechanisms of SNe IIb, contributing to our broader comprehension of stellar evolution and supernova diversity. 


Many Type IIb supernovae have a characteristic ``double-peak'' in their early light curves, presenting as a sharp rise within a few days of explosion followed by a rapid ($\sim$ hours-days) dimming, and then a gradual ($\sim$ weeks) re-brightening to a second peak before experiencing a final, long-lasting ($\sim$ months) decay \citep[e.g.,][]{Richmond1994,Arcavi2011,Kumar2013,Bufano2014,Morales-Garoffolo2014,Fremling2019,Armstrong2021,Pellegrino2023}. The secondary peak, similar in appearance and duration to those found in Type Ib/c SNe, is attributed to luminosity generated from the decay of nickel synthesized in the supernova explosion \citep{Arnett1980}. By contrast, the initial sharp rise to peak and subsequent rapid decay is attributed to the emission generated by the supernova shock heating and then cooling (by expansion) the extended envelope of the progenitor \citep{Soderberg2012}. \hl{Due to the short timescale and very early occurrence of the shock-cooling peak, it has only been observed in detail in a small number of previous objects \citep[e.g.,][]{Arcavi2017,Armstrong2021,Pellegrino2023,Wang2023}.}

Analytical calculations of the behavior of the extended progenitor envelope following deposition of the supernova energy predict a dependence \hl{on both the} properties of the progenitor and explosion \citep{Rabinak2011,Nakar2014}. This dependence makes analytical modeling of the shock-cooling light curve a powerful and effective probe for the stellar properties of the progenitor prior to explosion, even providing clues about the mass-loss history of the progenitor \citep[see e.g.,][]{Zimmerman2024}. The shock-cooling modeling approach has been used to create a small sample of progenitor estimates for supernovae on a case-by-case basis \citep[see e.g., ][]{Arcavi2017,Armstrong2021,Pellegrino2023}, but larger surveys with a consistent model application are needed to properly probe the space of progenitor channels and mass-loss modes.

In the last decade, multiple models for shock-cooling emission have been presented and tested on applicable supernova events. \hl{\cite{Piro2015} built on the earlier work of \cite{Nakar2014} to model the shock-cooling behavior pre- and post-peak}. \hl{\cite{Piro2021} then extended \cite{Piro2015} to include a two-zone model of variable power-law density in order to better model early shock-cooling emission behavior observed in several supernovae}. \hl{Unlike \cite{Piro2015}, which did not assume a specific polytropic structure for the stellar model, other approaches (e.g., \cite{Sapir2017}) assume a polytropic density profile and calibrate analytically derived expressions using numerical models.} \hl{A more recent model \citep{Morag2023} interpolated between the solutions of \cite{Sapir2017} and the previously developed \cite{SKW} model, resulting in an earlier validity than existing models}. \hll{As the model of \cite{Morag2023} is not calibrated specifically for Type IIb SNe and has never been applied to a Type IIb SN, the question of whether such a model can be applied to Type IIb SNe to take advantage of early-time data is of great interest.}

Here, we present early-time observations of the Type IIb SN 2022hnt coupled with a dense, multi-band, months-long follow-up campaign. We take advantage of the exceptionally early constraint on the shock-cooling rise of the light curve and infer the progenitor properties of the system using a variety of analytical shock-cooling models. Using SN 2022hnt as an example, we seek to present a framework for consistent shock-cooling model application to Type IIb SNe, taking into account: (i) regimes of validity, (ii) robustness of highly-constraining data points, (iii) temperature evolution uncertainties, (iv) intra- and inter-model variation, and (v) fiducial parameter construction. The paper is organized as follows. We provide a broad overview of our proposed modeling framework in \autoref{sec:overview_of_the_framework}. Next, we apply the framework to the new object, SN 2022hnt. We summarize the follow-up campaign and resulting data and reduction steps in \autoref{sec:data}. In \autoref{sec:evolution} we characterize the spectral evolution of the transient over the first five weeks. In \autoref{sec:fitting}, we characterize the light curve evolution and choose shock-cooling models from the literature to fit to the SN 2022hnt data. In \autoref{sec:physics}, we report the results of fits to the described models and compare and contrast the inferred progenitor properties both in the context of different model choices and different supernovae. Finally, in \autoref{sec:conclusions}, we summarize our results.

\section{Overview of the framework}
\label{sec:overview_of_the_framework}

\hl{Here, we propose a framework of considerations for the application of shock cooling models to Type IIb SNe. The goal of this framework is to provide a consistent procedure which maximizes the correctness and reliability of the resulting parameter estimates. The overall steps of this framework are:}
\begin{enumerate}
    \item [\textdagger\stepcounter{enumi}\arabic{enumi}.] Assess photometric and spectroscopic evolution of the transient,
    \item [\textdagger\stepcounter{enumi}\arabic{enumi}.]  Validate inference of shock cooling,
    \item [\textdagger\stepcounter{enumi}\arabic{enumi}.]  Select models for parameter estimation,
    \item [\textdagger\stepcounter{enumi}\arabic{enumi}.]  Compute and restrict data to regimes of model validity, 
    \item [\textdagger\stepcounter{enumi}\arabic{enumi}.]  Perform parameter estimates, varying model settings,
    \item [\textdagger\stepcounter{enumi}\arabic{enumi}.]  Reject estimates based on \textit{a priori} information, and
    \item [\textdagger\stepcounter{enumi}\arabic{enumi}.]  Construct fiducial estimates from non-rejected models.
\end{enumerate}
\hl{We expand on these items below.}

\vspace{0.3cm} 
\noindent \hl{\textit{Assess photometric and spectroscopic evolution of the transient}, \textdagger1: In general, the parameters we seek to estimate from models of shock cooling are: (i) the mass of the hydrogen-rich envelope, (ii) the initial velocity of the radiation-mediated shock, (iii) the radius of the progenitor star prior to explosion, and (iv) the explosion epoch. Properties of the transient evolution can be used to develop expectations for these parameters, which can in turn be used to reject models that are clearly incorrect based on these expectations. For example, \cite{Hiramatsu2021} showed that the hydrogen-rich envelope mass had an effect on the early light curve, with more hydrogen resulting in a flatter light curve (the ``short-plateau''). Additionally, the presence of non-detections and other tight \textit{a priori} constraints on the explosion epoch can provide a straightforward validation of fits.}

\vspace{0.3cm} 
\noindent \hl{\textit{Validate inference of shock cooling}, \textdagger2: Models of shock cooling are only applicable when the assumption of shock cooling is itself valid. Alternative light curve drivers (e.g., wind breakout or other heating) can produce similar photometric evolution while resulting in incorrect and misleading fits as a result of the invalid cooling assumption. During this step, photometry and spectra should be used to gauge the temperature evolution of the transient. Data for shock cooling modeling should be restricted exclusively to times when the envelope is cooling, and in particular where the temperature evolution falls off as $T\propto t^{-0.45}$, the nominal prediction of shock cooling \citep{Rabinak2011}. If the envelope is being heated e.g., due to a wind breakout, estimates and/or upper limits can be placed on the extent of the wind and the mass enclosed (see \cite{Zimmerman2024} for an overview, or \autoref{sec:validation_sce} for a demonstration on SN 2022hnt).}

\vspace{0.3cm} 
\noindent \hl{\textit{Select models for parameter estimation}, \textdagger3: There are multiple models of shock cooling available, divided into two classes: the Piro+ models \citep{Nakar2014,Piro2015,Piro2021} and the SW models \citep{Rabinak2011,SKW,Sapir2017,Morag2023}. The two model classes differ substantially in assumptions, implementation, and calibration; as a result, we recommend fitting at least one model from each class when possible for cross-model validation of parameter estimates. Generally, each model iteration improved upon the last, making it only necessary to fit the newest iteration. For example, \cite{Piro2021} extended \cite{Piro2015} from a one-zone to a two-zone model, while \cite{Morag2023} extended \cite{Sapir2017} to describe the spherical phase of the ejecta expansion. However, we recommend fitting previous model iterations if: (i) a model is so new that it has not been yet well established, making a comparison to previous iterations a useful sanity check, or (ii) when comparing parameter estimates to other objects in the literature which have used the previous model for analysis.}

\vspace{0.3cm} 
\noindent \hl{\textit{Compute and restrict data to regimes of model validity}, \textdagger4: Performing fits using a model that is not valid for particular data points will lead to erroneous and possibly misleading parameter estimates. Once a suite of models have been selected, the data selection must be tailored to the regimes of validity for each model. This consideration is particularly important when fitting a model which has been calibrated for one object class (e.g., Type II SNe) to another object class (e.g., Type IIb SNe). Some models provide analytic validity timescales which vary with the model parameters. These validity timescales should be implemented in the fitting routine, so that no iteration of the fit extends the model beyond its regimes of validity.}

\vspace{0.3cm} 
\noindent \hl{\textit{Perform parameter estimates, varying model settings}, \textdagger5: Many models have \textit{a priori} assumptions (e.g., polytropic index, emission suppression, etc.) which can be varied, resulting in different parameter estimates. 
The spread of values resulting from varying model assumptions will help characterize the true robustness of the final parameter estimates in a more comprehensive way than an uncertainty from a single fit would.}

\vspace{0.3cm} 
\noindent \hl{\textit{Reject estimates based on \textit{a priori} information}, \textdagger6: Using the expectations set in \textdagger1, incorrect or clearly erroneous model fits can be rejected to improve the accuracy and robustness of the fiducial parameter set. For example, a model which predicts an envelope mass $\gg 5 M_\odot$ can be discounted if the light curve shows a prominent double peak, based on the results of \cite[][Fig. 5]{Hiramatsu2021}. Similarly, fits which suggest a very early explosion epoch can be rejected if the estimated explosion epoch is incompatible with the light curve (e.g., there are multiple robust non-detections after the modeled explosion epoch).}

\vspace{0.3cm} 
\noindent \hl{\textit{Construct fiducial estimates from non-rejected models}, \textdagger7: The result of the previous steps is a curated set of parameter estimates from a variety of model classes and \textit{a priori} assumptions, probing both the uncertainty in the data and the uncertainties of the model assumptions. Construct a fiducial estimate using a method which incorporates both the variation of the parameter estimates and the cross-model parameter spread into a single estimate and uncertainty (e.g., a weighted mean and uncertainty).}
\section{Observations and data}
\label{sec:data}

\hl{SN 2022hnt was discovered at \textit{g}-band magnitude 18.1 by the Zwicky Transient Facility (ZTF) on April 14, 2022 \citep{22hntTNS}. Here, we present a summary of the follow-up campaign organized by ZTF, our own follow-up campaign using Las Cumbres Observatory (LCO) via the Global Supernova Project (GSP), early detections by ATLAS \citep[ATLAS;][]{Tonry2018}, and publicly-available ultraviolet (UVOT) observations from \textit{Swift} \citep{Swift}.}

\begin{figure*}
    \centering
    \includegraphics[scale=0.4]{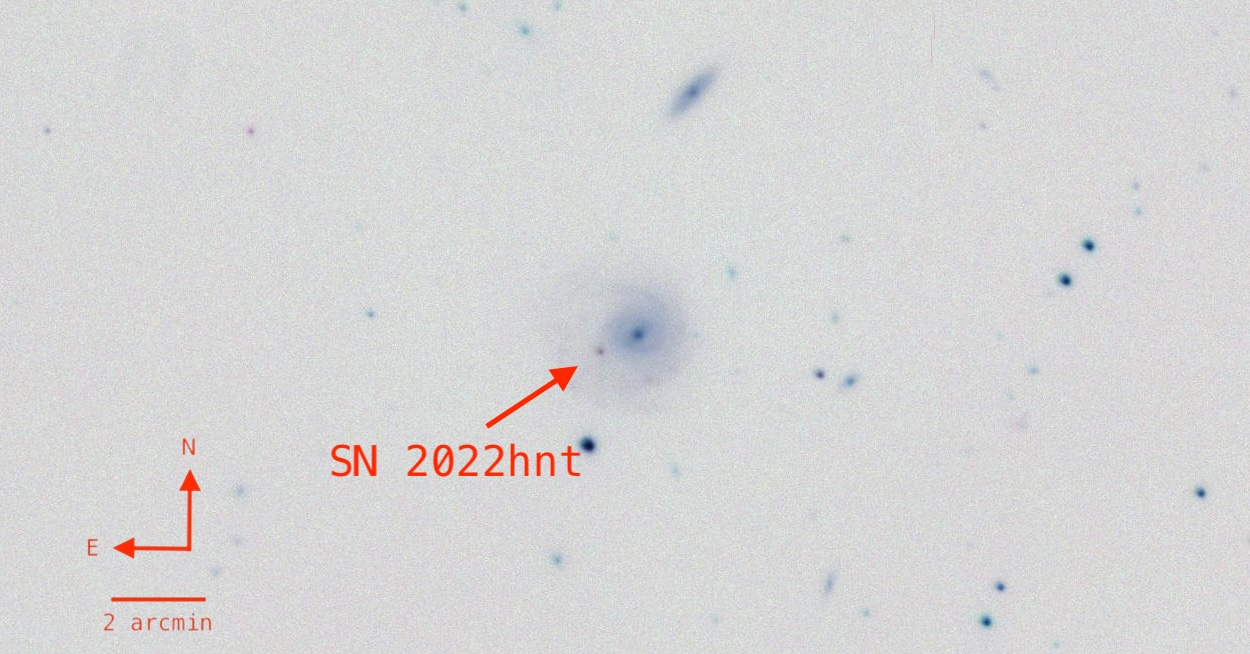}
    \caption{An image of SN 2022hnt $\sim$2 days after explosion. This image was made by stacking 80 minutes of LCO exposures in \textit{U, V, B, g, r} and $i$ bands. Colors are inverted to more clearly show the host and supernova. }
    \label{fig:sn2022hnt}
\end{figure*}

SN 2022hnt was detected at right ascension (R.A.) 11:36:59.751 and declination (decl) +55:09:50.26 in the young spiral galaxy NGC 3759A. An image of the object composed of stacked LCO exposures from $\sim2$ days post-explosion is shown in \autoref{fig:sn2022hnt}. The object was classified as a Type IIb SNe at redshift $z\sim0.0192$ in the Transient Name Server (TNS) using a spectrum obtained by the Spectral Energy Distribution Machine (SEDm) instrument mounted on the 1.5m Palomar telescope operated by ZTF \citep{22hntTNS}. Based on this redshift, and the distance to the host reported in the NASA Extragalactic Database (NED), we adopt a distance of 84 Mpc to the supernova. From the  \cite{Schlafly2011} dust map, we adopt an extinction along the line-of-sight to SN 2022hnt of $E(\mathrm{B-V})_{\mathrm{MW}}\sim0.0130$ mag. All light curve and photometric data presented in this paper are corrected using this extinction assumption. 

LCO is a global robotically-enabled network of telescopes ranging from 0.4m to 2m designed for rapid follow-up of active transients, as part of the GSP. We organized an intense follow-up campaign within 24 hours of discovery. Photometric observations in the $U$, $B$, $V$, $g$, $r$, and $i$ bands were made using the 1m and 2m telescopes in the network.
We reduced all LCO data using the \texttt{lcogtsnpipe} photometric reduction pipeline, which incorporates point-spread-function fitting and color term calculation prior to extraction and magnitude measurement \citep{Valenti2016}. We calibrated photometric observations in the $U$, $B$, and $V$ bands to the Landolt standard fields, which uses Vega magnitudes \citep{Landolt2009}. Observations in all other bands were calibrated to the Sloan Digital Sky Survey (SDSS) catalog, which uses AB magnitudes \citep{Smith2002}. 
Though the supernova was detected in close proximity to its host, the apparent magnitude of the host spiral arms is relatively low, and as a result, host subtraction was not required for the first 150 days of the observation. 

In addition to photometry, we also used LCO to monitor the spectroscopic evolution of SN 2022hnt. We obtained five spectra between day 5 and day 35 using the FLOYDS spectrograph mounted on the 2m telescope at Haleakala Observatory. The FLOYDS spectrograph covers wavelengths from 350 nm to 1 micron. To process the data, we employed the specialized LCO \texttt{floydsspec} pipeline, which handles cosmic ray flagging, spectrum extraction, and calibration \citep{FLOYDS}. The spectroscopic observations were used to estimate extinction due to the host, following the Na I D equivalent width method described in \citep{Poznanski2012}. We estimate host extinction of $E(\textrm{B}-\textrm{V})_{\mathrm{host}} \sim 0.0109$ mag. We apply this extinction correction globally to all photometry collected on SN 2022hnt.


As part of the LCO follow-up campaign, we triggered the Ultraviolet and Optical Telescope instrument on the \textit{Swift} observatory \citep{Swift}. We obtained several epochs of UV photometry spanning the first few weeks of the supernova evolution. The reduction of the ultraviolet photometry was performed using the \textit{Swift} Optical and Ultraviolet Supernova Archive (SOUSA) pipeline, with the most up-to-date zero points and calibrated sensitivity \citep{SwiftCal}.


ATLAS is a robotically-controlled set of two 0.5m survey telescopes located in Hawaii, on Mauna Loa and Haleakala \citep{Tonry2018}. ATLAS performed multiple observations of SN 2022hnt over a period of 90 days during the explosion, including a $\sim3\sigma$ detection prior to the first peak, which tightly constrains the rise time of the supernova. We discuss vetting of this detection in \autoref{sec:atlas_det}. The ATLAS photometry was obtained in the custom ATLAS ``orange'' ($o$-band) filter. Photometry was reduced using the proprietary ATLAS pipeline with calibration to the Pan-STARRS catalog. 

ZTF is a wide-field, nightly astronomical transient survey camera mounted on the 48-inch Palomar Observatory robotic telescope \citep{Bellm2019}. ZTF initially discovered the object and reported non-detections down to $g$-band 20th magnitude within hours of the first detection \citep{22hntTNS}. We incorporate the publicly available ZTF detections between April 2, 2022 and June 19, 2022 into our dataset. Subtractions are automatically performed using previous template images collected by the ZTF survey. Spectra for SN 2022hnt reported by ZTF were obtained via the SEDm on the 60-inch telescope at Palomar Observatory and downloaded from TNS. The SEDm is a low-resolution ($R\sim100$) spectrograph and reports flux on a wavelength range from $\sim350$ nm to $\sim900$ nm \citep{Masci2023}.

\begin{figure*}
    \centering
    \includegraphics[scale=0.75]{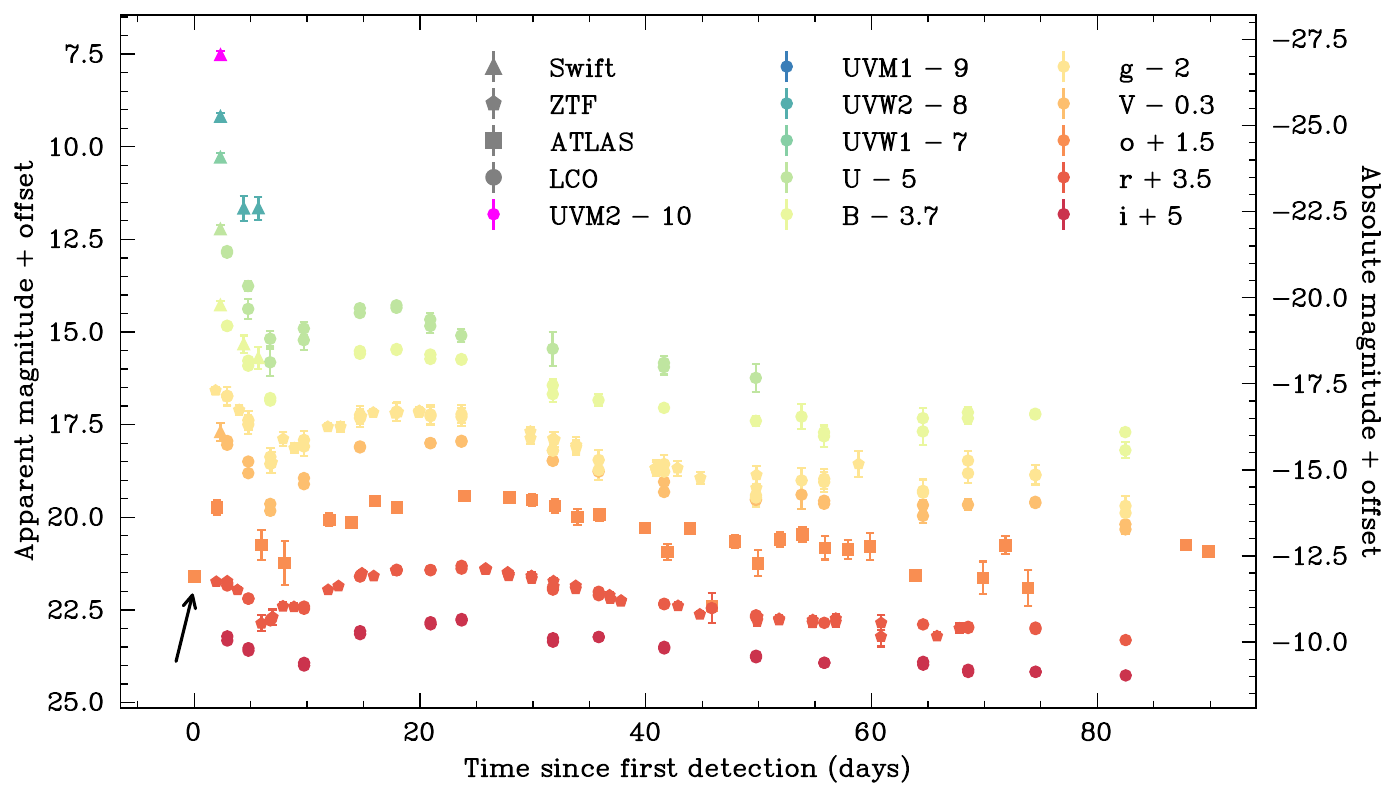}
    \caption{A compilation of the photometric data collected on SN 2022hnt by LCO, ZTF, \textit{Swift}, and ATLAS follow-up programs. Apparent and absolute magnitudes are shown offset arbitrarily to improve visibility. The ATLAS detection constraining the rise is indicated with an arrow. Error bars represent uncertainties of one standard deviation.}
    \label{sec:data:fig:compilation}
\end{figure*}

A compilation of all the photometry data described above is presented in \autoref{sec:data:fig:compilation}. A compilation of all spectra described above is presented in \autoref{fig:full_spectra}.

\subsection{Examination of the earliest ATLAS detection}
\label{sec:atlas_det}

The earliest ATLAS detection at $t_{\mathrm{AT}}=59681.4$ (``AT'' = ``ATLAS'') requires more scrutiny than the remainder of the dataset for two reasons: (i) the relatively low confidence of the detection and (ii) the impact of very early emission being probed with a robust detection. Unlike the remainder of the ATLAS detections, the highest-SNR raw detection reported by ATLAS at $t_{\mathrm{AT}}$ is a $2\sigma$ detection (measurement of $66\pm33 \ \mu$Jy). Given the proximity to the first unambiguous detections of the transient, the $t_{\mathrm{AT}}$ detection, if non-spurious, would play a significant role in constraining models of the light curve.

\begin{figure}
    \centering
    \hspace*{-1.cm}
    \includegraphics[scale=0.6]{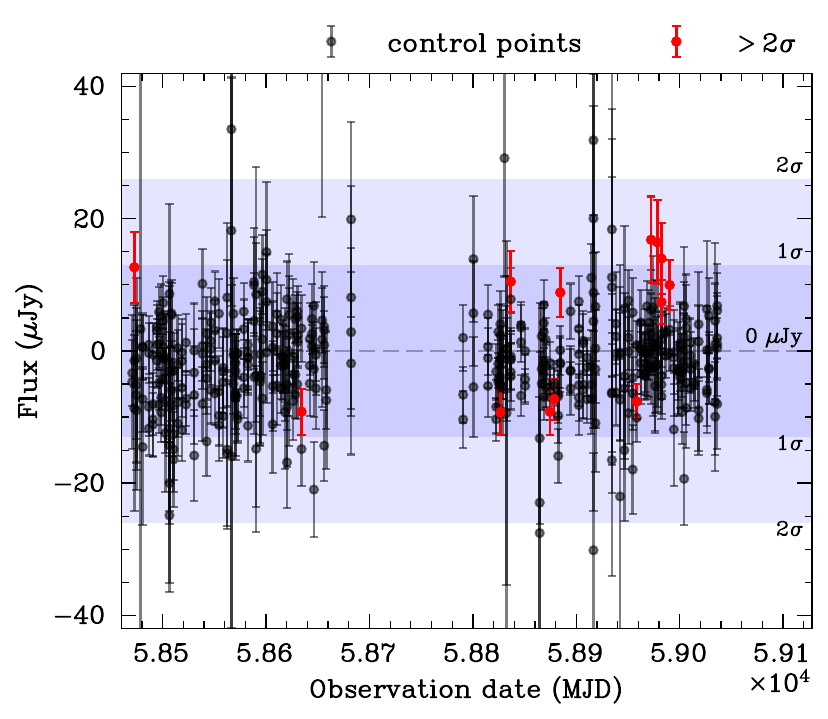}
    \caption{\hl{Example subset of the ATLAS control curves used to assess the validity of earliest ATLAS detection. The earliest ATLAS detection at $t_{\text{AT}} = 59861.4$ has a $2\sigma$ error. Examining the control curves, we find that $2\sigma$ detections (red) occur $<5\%$ of the time, an indication that the errorbar may be moderately conservative. The intrinsic scatter of the measurements ($\approx26$ $\mu$Jy, darker blue shaded region) is also lower than the $t_{\text{AT}}$ measurement errorbar ($33$ $\mu$Jy) would suggest. By co-adding all exposures at $t_{\text{AT}}$ and adopting the $26$ $\mu$Jy errorbar, we can increase the significance of the $t_{\text{AT}}$ detection to $2.7\sigma$.}}
    \label{fig:atlas_control}
\end{figure}

\hl{The observations reported by ATLAS at $t_{\mathrm{AT}}$ consist of three exposures, acquired within 15 minutes of each other. Two exposures are $\sim1\sigma$ and the third is an exactly $2\sigma$ detection. To evaluate, we use the tool \texttt{ATclean} \citep{ATClean} to perform cuts based on eight control epochs, drawn from a circular region around the transient location \citep[see][for further details on the procedure]{ATClean}. An example subset of the control epochs is shown in \autoref{fig:atlas_control}. All three data points exceed the minimum constraints required to be considered real; namely, all three points have uncertainty $<160\ \mu$Jy and a PSF fit $\chi^2 < 20$. }

Next, we seek to improve our understanding of the emission captured at $t_{\mathrm{AT}}$. We stack all three intranight exposures at this epoch using the given uncertainties and produce a co-added measure of the photometry, which results in a slight increase of the confidence from $2\sigma$ to $2.1\sigma$. However, we suggest that the ATLAS error bar may be a mild overestimate of the uncertainty present in the measurement. Examining the no-flux control curves after applying the cuts described above, we note that $2\sigma$ measurements appear $\ll5\%$ of the time, less than would be expected for a true $2\sigma$ measurement. Additionally, the spread associated with no flux in the control curves is $\sim\pm26 \ \mu$Jy, smaller than the errors reported by all three observations acquired at $t_{\mathrm{AT}}$. \hl{Both these effects are visualized using a subset of the control curves in \autoref{fig:atlas_control}}. If we apply this estimate of the uncertainty to the observations at $t_{\mathrm{AT}}$, it expands the co-added significance to $\sim2.7\sigma$. While this confidence does not make the measurement highly robust, we choose to include it with caution in the analysis based on the above discussion and the proximity of the detection to the transient. 

\section{Evolution of the transient}
\label{sec:evolution}

\subsection{Spectral evolution}
\begin{figure*}
    \centering
    \includegraphics[scale=0.7]{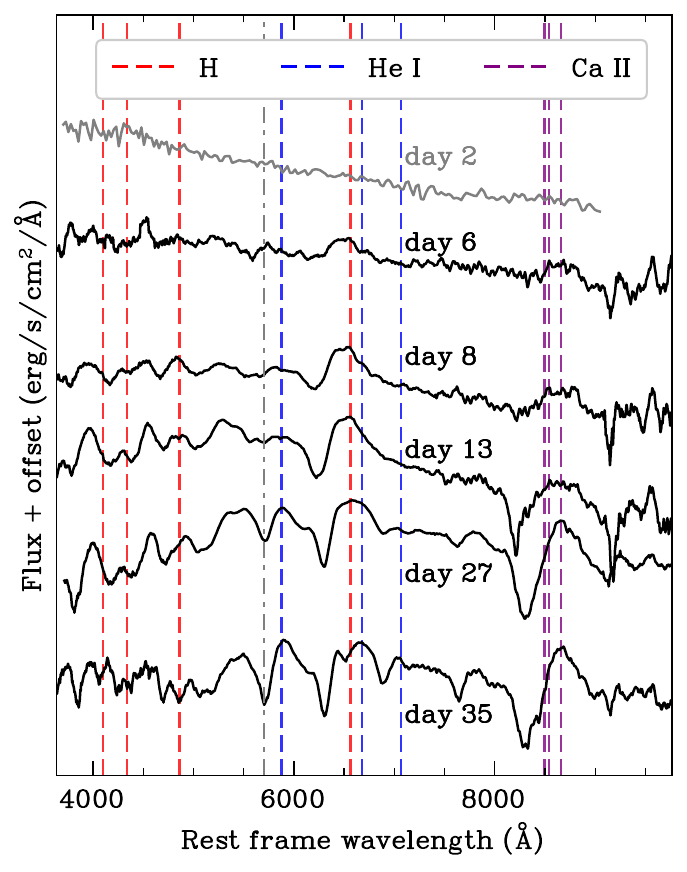}
    \includegraphics[scale=0.695]{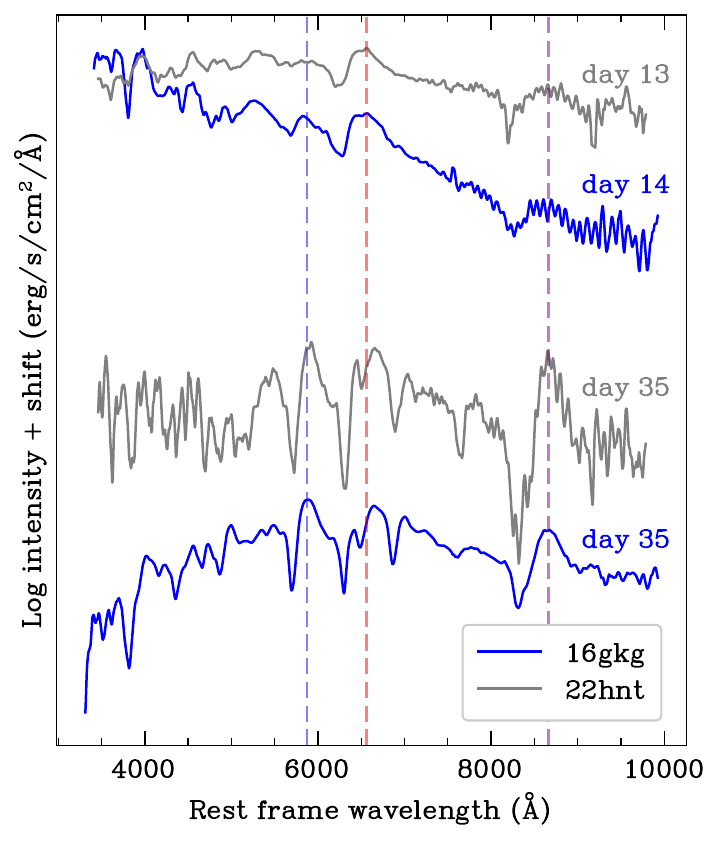}
    \caption{(\textit{left}) The complete spectral evolution of SN 2022hnt (LCO, black; ZTF, gray) in rest wavelength. We highlight the following spectral lines as dashed vertical lines: hydrogen (red), helium (blue), and Calcium II (purple). The gray dot-dashed line at $\sim5700$ \angstrom shows He I blueshifted by $\sim8790$ km/s. (\textit{right}) A comparison of the spectra of 22hnt to another well-known IIb, SN 2016gkg, at two representative phases. Dashed lines show the location of rest-frame hydrogen (red), helium (blue), and calcium (purple). Both spectra are shown in their rest frame wavelength. The spectrum of SN 2022hnt is quite consistent with other objects in the Type IIb subclass.}
    \label{fig:full_spectra}
\end{figure*}

The earliest LCO spectrum of SN 2022hnt was taken within one week of the estimated explosion date. Over the next $\sim5$ weeks, we recorded a spectrum of the transient once every $\sim1$ week, for a total of 5 spectra covering the first month of the evolution of the supernova. An additional ZTF spectrum was taken during the first 24 hours post-discovery \citep{22hntTNS}. The complete spectral evolution of SN 2022hnt (LCO+ZTF) is shown in \autoref{fig:full_spectra}. The spectral evolution is broadly consistent at representative phases with other Type IIb SNe such as SN 2016gkg, also shown in \autoref{fig:full_spectra}. 

The earliest spectrum (taken by ZTF, obtained from TNS) show a largely featureless blackbody, typical for a IIb supernova in the earliest explosion stage. The high temperature blue spectrum is consistent with a very hot fireball; spectral lines will not begin to show until the ejecta have cooled down considerably \citep{Filippenko1997}. The spectrum is too low-resolution to conclusively rule out flash features indicating interaction with a circumstellar medium (CSM) as are sometimes found in Type II SNe in the few hours post-explosion \citep[e.g.,][]{Flash1, Flash2}. We do not note any such features in the SN 2022hnt spectra. 

The next two spectra correspond to $\sim1$ week post-explosion. At this stage, the ejecta have shock cooled and several lines are beginning to develop. We identify the presence of a broad H$\alpha$ bump in the day 6 spectrum which develops into a clear P Cygni line profile as the line-of-sight absorption becomes more prominent. Additionally, we note the presence of Ca II, definitively by day 8 at 4000 and 9000 \angstrom, and possibly but not definitively by day 6. 

By day 13, the H$\alpha$ has strengthened and the absorption feature is more clear, almost equal in magnitude to the emission feature. A Mg I feature is clearly visible at $\sim5500$ \angstrom. This feature may have been present at day 8 as a weak bump, but is a prominent emission line by day 13. Both the blue and red Ca II lines have strengthened as well, particularly the narrower Ca II doublet at $\sim$ 4000 \angstrom. The H$\alpha$ emission is continuing to increase in intensity but is flattened by the presence of an increasingly blueshifted He I P Cygni profile, as is common for Type IIb SNe \citep{Filippenko1997,2000AJ....120.1487M,Pellegrino2023}. 

By day 27, the H$\alpha$ P Cygni, Ca II lines, and Mg I are at their maximum intensities achieved during our monitoring. We note blueshifted He I absorption features coincident with the H$\alpha$ emission as well at $\sim5800$ \angstrom. The blueshifted helium emission originates from a more interior layer of the ejecta and is an effective approximation of initial shock velocity in the absence of emission profiles for more interior elements (e.g., Fe III). To estimate the velocity of the He I absorption lines, we fit a Gaussian profile to each feature and interpreted the best-fit mean as the location of the feature. 
Though the absorption line at $\sim5800$ \angstrom is potentially coincident with a nearby Na line, we noted that both the absorption feature at $\sim5800$ \angstrom and the He I absorption line coincident with H$\alpha$ evolved similarly, indicating either would be a suitable choice to measure the He I velocity. 
We find a corresponding line blueshift of $\sim8790$ km/s on both the stronger absorption feature at $\sim5800$ \angstrom and the absorption feature coincident with the H$\alpha$ P Cygni profile. We perform this fit for all days on the stronger feature when visible and find a slight decrease in velocity from $\sim9000$ km/s to $\sim7000$ km/s over the first 35 days of the explosion. 

Following 27 days post-explosion, most emission and absorption features begin to decrease in intensity, particularly the H$\alpha$ feature, as is common for a hydrogen-stripped IIb \citep{Filippenko1997}.

\subsection{Photometric evolution}

The full photometric evolution of SN 2022hnt is shown in \autoref{sec:data:fig:compilation}. This visualization includes all LCO photometry as well as photometry obtained via ATLAS, ZTF, and Swift. The early ($t<1$ week) luminosity is almost entirely dominated by the emission from shock cooling as the shocked envelope expands \citep{Soderberg2012}. During these epochs, the emission is blue and rapidly diminishing. The rise to the first peak is constrained by an early sequence of serendipitous ATLAS detections immediately prior to the peak of the shock cooling emission. Due to the rise to peak being only ${\sim}$hours in duration, it is extraordinarily difficult to organize and execute follow-up campaigns which probe the behavior during this phase. Thus, the presence of rigorous detections on the rise represents an opportunity to test the effectiveness of existing shock cooling models at describing behavior during this phase.

After the shock cooling phase, the emission in the explosion is largely dominated by the decay of $^{56}$Ni synthesized during the explosion into $^{56}$Co. The onset of this emission causes a slower re-brightening of the object to a second peak $\sim1$ month into the evolution of the transient \citep{Arnett1980}. The remainder of the light curve is dominated by this $^{56}$Ni decay and further decays from $^{56}\mathrm{Co}\to$$^{56}$Fe, which forms an elongated ``tail''.

In the following sections, we fit a series of physically-motivated models to the shock-cooling phase of SN 2022hnt. These models use the light curve behavior to directly measure physical properties of the explosion, progenitor, and ejecta. 
\section{Photometric modeling}
\label{sec:fitting}

In order to extract physical measurements from the data, we fit several physically motivated, analytic models of the shock-cooling light curve behavior. Of interest is assessing how the various models agree and conflict, which can identify limitations and inform which models may be most consistent for making measurements on a sample of Type IIb supernovae. Previous analyses reported significant disagreements in shock cooling parameter estimates between different models \citep{Pellegrino2023}, as well as applications of the same model with different initial assumptions \citep[e.g., polytropic index; see e.g.,][]{Arcavi2017}. Additionally, shock cooling models must be applied with careful attention to regimes of validity in order to report conservative and accurate results \citep[see e.g.,][for a discussion]{Zimmerman2024}.

To address these problems, we propose a framework for incorporating the most impactful considerations of shock cooling into light curve modeling attempts, with the goal of producing the most robust parameter estimates possible (see \autoref{sec:overview_of_the_framework} for a full description). In \autoref{sec:validation_sce}, we discuss validation of the shock cooling assumption and argue for the placement of upper limits on the configuration of a heating scenario if cooling cannot be confirmed back to the explosion epoch.  In \autoref{sec:fitting:subsec:sw17}, \autoref{sec:fitting:subsec:p21}, and \autoref{sec:fitting:subsec:m23}, we present the necessary information to fit several models of shock cooling, with the intention of using consistency between the models to validate parameter inference. In \autoref{sec:fitting:subsec:regimes}, we consider the mass, temperature, and transparency regimes of validity for each model, which will be used to caveat the reliability of fiducial parameter estimates. Finally, in \autoref{sec:fitting:subsec:procedure}, we describe our fitting procedure and motivate variation of pre-fitting considerations (e.g., model assumptions, selective data exclusion, etc.), to better understand the robustness of the parameter inference within the context of the individual models.

\subsection{Validation of shock cooling inference}
\label{sec:validation_sce}

\begin{figure}
    \centering
    \hspace*{-0.7cm}\includegraphics[scale=0.85]{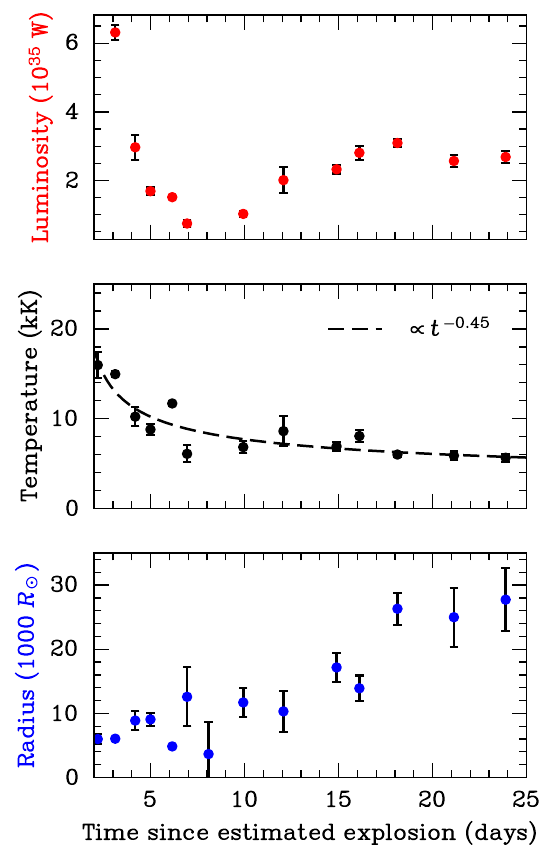}
    \caption{Blackbody fits of temperature, radius, and luminosity to the multicolor data of SN 2022hnt. Radial expansion is consistent with a linear coasting phase; temperature shows a decline consistent with the nominal prediction of shock cooling, a $T\propto t^{-0.45}$ power law. The luminosity fits clearly reproduce the double peak characteristic of Type IIb SNe. }
    \label{sec:fitting:fig:blackbody}
\end{figure}

The choice to fit models of shock cooling to the early time emission must first be validated by ensuring that cooling is in fact taking place. Heating mechanisms (e.g., wind breakout, where the shock breaks out of an extended CSM beyond the stellar surface) can produce similar light curves to shock cooling but will produce erroneous fits due to the underlying incorrect cooling assumption. We investigate the cooling assumption by performing blackbody luminosity, temperature, and radius fits for each epoch with multicolor photometry. The results of the blackbody fits are shown in \autoref{sec:fitting:fig:blackbody}. The bolometric luminosity shows a clear double peak, characteristic of the Type IIb subclass. The blackbody temperature shows monotonic decay well-approximated by a $t^{-0.45}$ power-law, the nominal theoretical prediction of shock cooling \citep{Rabinak2011}. Finally, the radius grows linearly with time, consistent with expected behavior as the supernova is in the coasting phase of its expansion. The result of these fits are consistent with a shock cooling phenomenon occurring for least $t\gtrsim2$ days, corresponding to the second epoch of data and beyond.

The assessment of shock cooling, while valid for all data points $t\gtrsim2$ days, can not be as easily made for the earliest data epoch ($\sim$59681.3 MJD). Due to the spectrum of the supernova likely peaking at UV wavelengths, a single $o$-band measurement is insufficient to constrain the temperature at this epoch. As a result, we cannot definitively determine the temperature gradient at $t<2$ days, which would conclusively confirm or rule out shock cooling. The worst-case scenario, where the object temperature evolution is consistent with heating due to wind breakout until $t\sim2$ days, places an upper limit on the extent of the CSM $R_{\textrm{bo}}$ being heated by the shock. Following \cite{Zimmerman2024}, we use the relation $R_{\textrm{bo}} \sim t_{\textrm{rise}} v_s$, where $t_{\textrm{rise}} \lesssim 2$ days is the maximum possible rise time, and $v_s$ is the estimated shock velocity. We estimate an upper limit on the extent of the CSM to be
\begin{equation}
    R_{\textrm{bo}} \lesssim 2500 \Bigg(\frac{t_{\textrm{rise}}}{2 \textrm{ days}}\Bigg)\Bigg(\frac{v_s}{10^4 \textrm{ km/s}}\Bigg) R_\odot.
    \label{eq:windbreakout_r}
\end{equation}
This estimate for $R_{\textrm{bo}}$ is consistent with (i.e., less than) the earliest blackbody radius fits,  $R_{\textrm{bo}} \lesssim 7000 R_\odot$. We can similarly upper bound the mass of the shocked CSM emitting prior to breakout from the extended wind. Again following \cite{Zimmerman2024}, we approximate the characteristic bolometric luminosity at breakout $L_{\textrm{bo}}$ as the total amount of kinetic energy deposited into the CSM, $L_{\textrm{bo}} \sim Mv_s^2/t_{\textrm{rise}}$, where $M$ is the mass of the CSM interior to $R_{\textrm{bo}}$. We estimate an upper limit on the mass contained as,
\begin{equation}
    M \lesssim 8.7\Bigg(\frac{L_{\textrm{bo}}}{10^{36} \textrm{ watts}}\Bigg)\Bigg(\frac{t_{\textrm{rise}}}{2 \textrm{ days}}\Bigg) \Bigg(\frac{v_s}{10^4 \textrm{ km/s}}\Bigg)^{-2} 10^{-4}  M_\odot.
    \label{eq:windbreakout_m}
\end{equation}
By simply adopting $R_{\textrm{bo}}$ into a volume, we can estimate a characteristic CSM density of $M/R_{\textrm{bo}}^3 \sim 3.3\times10^{-13} \mathrm{\ g\ cm^{-3}}$.

\subsection{Choice of shock cooling model}
\subsubsection{\cite{Sapir2017} Model}
\label{sec:fitting:subsec:sw17}

\cite{Sapir2017} addressed the emission generated during $t\lesssim$ a few days post-explosion (when the luminosity is primarily generated in the layers of the ejecta proximate to the stellar surface), but also extended the model of \cite{Rabinak2011} to $t\gtrsim$ a few days post-explosion, when the luminosity is generated in a $\rho$-profile dependent manner from the innermost layers of the material. 

The luminosity and resulting color temperature is dependent on the assumed polytropic index $n$, which itself varies depending on the assumed physical situation. For typical red supergiant stars with convective envelopes, we assume a polytropic index of $n=3/2$. However, in stellar environments dominated by a radiative envelope (e.g., blue supergiants), a polytropic index of $n=3$ is more appropriate. We modify the final expression for the luminosity from \cite{Sapir2017} to account for choice of polytropic index $n$ and obtain
\begin{align}
    L(t) &\sim (1.88-0.147[n-3/2])\times10^{42} \times \nonumber \\
    & \left(\frac{v_{s,8.5} t^2}{f_\rho M \kappa_{0.34}}\right)^{3\times10^{-3}-5.9\times10^{-2}n} \times \nonumber \frac{v_{s, 8.5}^2 R_{13}}{\kappa_{0.34}}\times \nonumber \\
    & \exp\left(-\frac{\left[1.93nt-1.23t\right]}{\left[19.5\kappa_{0.34} M_e v_{s,8.5}^{-1}\right]^{0.5}}\right)^{0.87-4.7\times10^{-2}n} \mathrm{\ erg \ s^{-1}},
    \label{eq:lt_sw17}
\end{align}
where $v_{s,8.5}$ is the velocity of the shock in units of $10^{8.5} \mathrm{\ cm \ s^{-1}}$, $t$ is the phase of the supernova in days, $M$ is the total mass of the star (core $M_c$ plus envelope $M_e$) in solar masses, $\kappa_{0.34}$ is the opacity in units of $0.34 \mathrm{\ cm^2 \ g^{-1}}$, $R_{13}$ is the radius of the extended envelope of the progenitor star $R_e$ in units of $10^{13}$ cm, and
\begin{equation}
f_\rho \sim \begin{cases}\left(M_e / M_c\right)^{0.5} & n=3 / 2 \\ 0.08\left(M_e / M_c\right) & n=3\end{cases}.
\end{equation}
In order to fit multiband data such as that collected for SN 2022hnt, \cite{Sapir2017} also derived a color temperature for the expanding photosphere as a function of time. We transform the temperature expression similarly to \autoref{eq:lt_sw17} and obtain
\begin{align}
    T(t) &\sim (2.14-0.06n) \times10^4 \times \nonumber \\
    & \left(\frac{v_{s,8.5}^2 t^2}{f_\rho M \kappa_{0.34}}\right)^{3.8\times10^{-2} - 7.3\times10^{-3}n} \times \frac{R^{0.25}_{13}}{\kappa_{0.34}^{0.25}}t^{-0.5}\mathrm{\ K}.
    \label{sec:fitting:eq:sw17_t}
\end{align}

\subsubsection{\cite{Piro2021} Model}
\label{sec:fitting:subsec:p21}

\hl{The basic model of \cite{Piro2015} was revisited and extended in \cite{Piro2021}. The extended model incorporated low-mass extended material into a two-zone model with a broken power law dependence. The two-zone model consists of an outer region with a steep radial dependence for density and an inner region with a shallow radial dependence. The behavior was characterized before and after a time $t \approx t_d$, corresponding to the approximate time the diffusion reaches the boundary between the inner and outer zones. We reproduce the analytic expression for the luminosity $t\lesssim t_d$ as calculated in \cite{Piro2021} here:}
\begin{equation}
    L(t) \approx \frac{\pi(n-1)}{3(n-5)} \frac{c R_e v_t^2}{\kappa}\left(\frac{t_d}{t}\right)^{4 /(n-2)}.
\end{equation}
\hl{After the diffusion reaches the zone boundary (i.e., $t\gtrsim t_d$) the luminosity was derived by \cite{Piro2021} to be}
\begin{equation}
    L(t)=\Upsilon \exp \left[-\frac{1}{2}\left(\frac{t^2}{t_d^2}-1\right)\right].
\end{equation}
\hl{In both of these expressions, $n\approx10$ is the steep power-law dependence of the outer zone, $R_e$ is the radius of the extended envelope, $\kappa$ is the electron scattering opacity, $t$ is the phase of the supernova in days, $\Upsilon$ is a prefactor defined as}
\begin{equation}
    \Upsilon=\frac{\pi(n-1)}{3(n-5)} \frac{c R_e v_t^2}{\kappa},
\end{equation}
\hl{and $v_t$ is the velocity at the zone boundary, defined as}
\begin{equation}
    v_t=\left[\frac{(n-5)(5-\delta)}{(n-3)(3-\delta)}\right]^{1 / 2}\left(\frac{2 E_e}{M_e}\right)^{1 / 2},
\end{equation} 
\hl{where $M_e$ is the mass of the extended envelope, $\delta\approx0.1$ is the shallow radial dependence of the inner zone, and $E_e$ is the fraction of the total energy of the supernova deposited in the extended material. We follow \cite{Piro2021} and choose}
\begin{equation}
E_e \approx 2 \times 10^{49} E_{51}\left(\frac{M_c}{3 M_{\odot}}\right)^{-0.7}\left(\frac{M_e}{0.01 M_{\odot}}\right)^{0.7} \mathrm{erg},
\end{equation}
\hl{Given the luminosity derived above the assumed radius, we use the Stefan-Boltzmann law and derive the following expression for the temperature as a function of supernova phase}
\begin{equation}
    T(t)=\left[\frac{L(t)}{4 \pi R^2(t) \sigma_{\textrm{SB}}}\right]^{1 / 4},
    \label{eq:blackbody_temp}
\end{equation}
\hl{where $\sigma_{\textrm{SB}}$ is the Stefan-Boltzmann constant.}

\subsubsection{\cite{Morag2023} Model}
\label{sec:fitting:subsec:m23}

\cite{Morag2023} treats the expansion of the supernova ejecta in two phases: (i) a planar phase where the width of the emitting region is
$ \delta r\ll R$ and is located at $r\sim R$, and (ii) a later, spherical phase
where the emitting region is located at $r\gg R$. Previous works  have presented exact and approximate solutions for the luminosity and color temperature in the planar and spherical phase, respectively \citep{Rabinak2011,
Katz2012}. \cite{Morag2023} presents an analytically-derived, numerically-calibrated interpolation between the exact solution previously derived for the planar phase and the approximate solution derived in \cite{Sapir2017} for the spherical phase of the expansion. 

We reproduce the interpolation here. For simplicity, the quantities in the interpolation are normalized to the time $t_{\textrm{br}}$ corresponding to the transition point between the planar and the spherical phase. \cite{Morag2023} reported the luminosity for the planar phase as
\begin{equation}
    L / L_{\mathrm{br}}=\tilde{t}^{-4 / 3}+A \exp \left[-\left(a t / t_{\mathrm{tr}}\right)^\alpha\right] \tilde{t}^{-0.17},
\end{equation}
where $\tilde{t}\equiv t/t_{\textrm{br}}$, $A\sim0.9$, $a\sim2$, $\alpha\sim0.5$. Additionally, 
\begin{align}
t_{\mathrm{tr}} & =\sqrt{\frac{\kappa M_{\mathrm{env}}}{8 \pi c v_{\mathrm{s} *}}} \\
& =19.5 \sqrt{M_{\mathrm{env}} \kappa_{0.34} v_{\mathrm{s} *, 8.5}^{-1}} \mathrm{~d},
\end{align}
where $v_{s\ast}\sim 1.05f_\rho^{-0.19} \sqrt{E/M}$ is a derivative quantity of the shock velocity, $M_{\mathrm{env}}=M_0-M_c$ is the envelope mass in solar masses, and
\begin{equation}
    L_{\mathrm{br}}=3.69 \times 10^{42} R_{13}^{0.78} v_{\mathrm{s} *, 8.5}^{2.11}\left(f_\rho M_0\right)^{0.11} \kappa_{0.34}^{-0.89} \mathrm{erg} \mathrm{s}^{-1}.
\end{equation}

In order to compare this to the multiband data collected on SN 2022hnt, we cite the corresponding normalized color temperature relation from \cite{Morag2023}:
\begin{equation}
    T_{\mathrm{col}} / T_{\mathrm{col}, \mathrm{br}}=\min \left[0.98 v_{\mathrm{bo}, 9}^{-0.05} \tilde{t}^{-1 / 3}, \tilde{t}^{-0.45}\right] ,
\end{equation}
where 
\begin{equation}
    v_{\text {bo }}=3.31 M_0^{0.13} v_{*, 8.5}^{0.13} R_{13}^{-0.26} \kappa_{0.34}^{0.13} f_\rho^{-0.09}\sqrt{E/M}.
\end{equation}
With the modification of the $t_{\textrm{tr}}$ quantity, the validity of this model was extended to $t\lesssim t_{\textrm{tr}}/a$. The model is more broadly valid in the shock-cooling phase up to hydrogen recombination, i.e., until the temperature in the ejecta drops to $\sim 0.7$ eV. \hll{However, this model is not numerically calibrated specifically for Type IIb SNe and has never been applied to Type IIb SNe. As a result, while the model may reproduce the light curve behavior to a reasonable degree, the anticipated performance of the model is uncertain and it is expected to be far less discriminatory than if it were applied to ordinary Type II SNe.}

\subsection{Regimes of validity}
\label{sec:fitting:subsec:regimes}

We consider regimes of validity for shock cooling models as applied to SNe Type IIb, which have lower envelope masses than the SNe Type II such models are normally calibrated against. We first consider the temperature regime of validity. Both \cite{Sapir2017} and \cite{Morag2023} are only valid for $T \gtrsim 0.7 \textrm{ eV} \sim 8120$ K. \hl{To ensure all fits are performed at valid temperatures, we restrict the data to epochs where the blackbody fits performed in \autoref{sec:validation_sce} reported a bolometric temperature $T > 8120$ K.} A rigid application of this rule (using a strict cutoff based on the blackbody fit values) restricts the data to $t \lesssim 7$ days. A less rigid application (using a looser cutoff based on the $t^{-0.45}$ power law fit in \autoref{sec:fitting:fig:blackbody}) restricts the data $t \lesssim 10$ days.

Next, we consider the transparency regime of validity. Shock cooling models incorporate an approximate late-time bolometric luminosity suppression factor, which is valid up to the time $t_{\textrm{tr}}/a$, when the envelope becomes transparent (i.e., optical depth $\tau\sim c/v$). The approximation for \cite{Sapir2017} [\cite{Morag2023}] is valid for mass ranges $0.1 M_\odot \lesssim M_{\textrm{env}} \lesssim 10 M_\odot$ [$1  M_\odot\lesssim M_{\textrm{env}} \lesssim 10 M_\odot$]. For the \cite{Sapir2017} model, this validity range will make the approximation applicable for a substantial range of IIb envelope masses, which typically live between $0.01 \ M_\odot \lesssim M_{\textrm{env}} \lesssim 0.5 \ M_\odot$. However, we are beyond the calibration regime for the model of \cite{Morag2023}, which creates an opportunity to assess the effectiveness of the calibration for Type IIb SNe. Assuming a valid or near-valid mass estimate, the transparency timescale $t_{\textrm{tr}}$ can be computed using equation (20) from \cite{Sapir2017} (or \cite{Morag2023} equation A3) as
\begin{equation}
    t_{\textrm{tr}} \sim 2 \Bigg(\frac{M_{\textrm{env}}}{0.1 M_\odot}\Bigg)^{1/2}\Bigg(\frac{v_s}{3\times10^{8.5} \mathrm{\ cm/s}}\Bigg)^{-1/2} \textrm{ days}.
    \label{eq:transparency_timescale}
\end{equation}
As this timescale is sensitive to the envelope mass and shock velocity (which will be varied throughout fitting), we dynamically implement the data cutoff by restricting the likelihood calculation for a given model to data with $t \lesssim t_{\textrm{tr}}/a + 0.5$ day. Photometric data were sampled on $\sim1$ day cadence; thus, the additional 0.5 day overhead allows photometry just outside the validity timescale to be included with minimal complications for the fit.

\begin{figure}
    \centering
    \hspace*{-0.7cm}\includegraphics[scale=0.6]{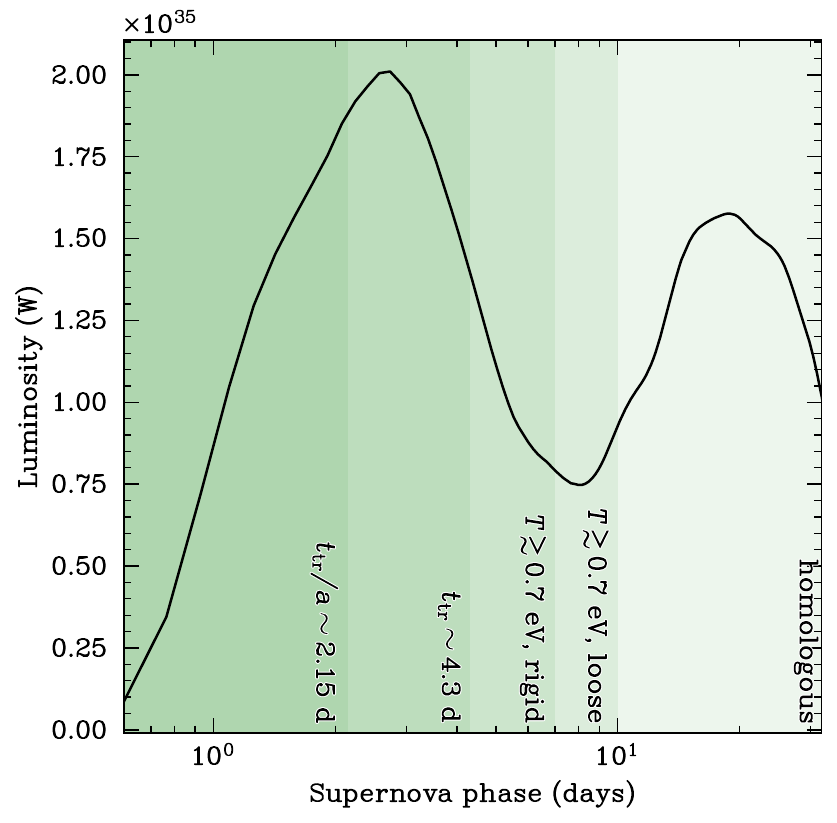}
    \caption{\hl{A visualization of how the regimes of model validity restrict fitting to subsets of the data for a model with $M=0.2 \ M_\odot, R=100 R_\odot$, and $v_s=5000$ km/s. The light curve is a smoothed fit to the luminosities of SN 2022hnt. From right to left, the restrictions are: (i) ``homologous'', a restriction to the part of the supernova evolution where the expansion can be considered homologous (effectively $t\gtrsim 1$ hr), (ii) ``$T\gtrsim0.7$ eV, loose'', a loose restriction (see \autoref{sec:fitting:subsec:regimes}) to the part of the supernova evolution where the temperature is high enough to facilitate hydrogen recombination, (iii) ``$T\gtrsim0.7$ eV, rigid'', the same as (ii) but a rigid restriction (see \autoref{sec:fitting:subsec:regimes}), (iv) ``$t_{\text{tr}}$'', a restriction to times less than the transparency timescale defined in \autoref{eq:transparency_timescale}, and (v) ``$t_{\text{tr}}/a$'', the most conservative restriction to the subset of the evolution with phase less than the scaled transparency timescale as described in \autoref{sec:fitting:subsec:regimes}.}}
    \label{sec:fitting:fig:validity}
\end{figure}

Finally, we consider the homologous expansion break time. \cite{Morag2023} interpolates between solutions for the planar and spherical phases (switching at time $t=t_{\textrm{br}}$, while \cite{Sapir2017} considers only the spherical phase. An expression for the break time is given by equation (40) of \cite{Morag2023},
\begin{equation}
    t_{\textrm{br}} = 0.86 \Bigg(\frac{R}{10^{13} \textrm{ cm}}\Bigg)^{1.26}\Bigg(\frac{v_{s*}}{10^{8.5} \textrm{ cm}}\Bigg)^{-1.13} (f_\rho M_0 \kappa_{0.34}) \textrm{ hrs}.
\end{equation}
The break time is not a validity timescale, but it is a useful quantity for comparing the models of \cite{Sapir2017} and \cite{Morag2023}. Provided $t_{\textrm{br}}$ is substantially earlier than the first epoch of data used in the fits, \cite{Sapir2017} and \cite{Morag2023} should report similar results. Thus, the homologous expansion break timescale forms the basis of a useful sanity check on the fits. A visualization of each of these validity ranges is shown in \autoref{sec:fitting:fig:validity}.

\subsection{Fitting procedure and considerations}
\label{sec:fitting:subsec:procedure}

We fit the models above via the open-source package \texttt{lightcurve-fitting} \citep{griffin_fitting}. \hl{This package implements a Markov-Chain Monte Carlo (MCMC) fitting routine, which we initialized with 15 walkers with $\gtrsim10^5$ steps ($10^3$-step burn-in) for posterior exploration. Fit settings (walkers, steps, etc.) were chosen based on two metrics of convergence: (i) the improved Gelman-Rubin statistics (specifically, $\hat{R}<1.01$) \citep{ImprovedRhat}, and (ii) chain autocorrelation (specifically, checking that the chain length is a factor of $\gtrsim50$ longer than the autocorrelation timescale) \citep{emcee}}. \texttt{lightcurve-fitting} performs the maximization and exploration of the resulting posterior using a wrapper of the MCMC algorithm implemented in the \texttt{python} package \texttt{emcee} \citep{emcee}. 
All data are weighted uniformly, with no de-emphasis based on distance from explosion date as conducted by, e.g., \cite{Arcavi2017}. \hl{We condition the model on the dataset using a simple Gaussian likelihood function. We investigated potential likelihood function biases (e.g., intrinsic scatter terms \citep{Apian}, quantifying deviations from a blackbody \citep{Irani2024}, etc.) and did not find these altered the results.} \hl{Confidence intervals on the maximum a posteriori value were determined from the posteriors by computing the highest posterior density interval \citep[as computed in, e.g.,][]{HPDI}, as all posterior distributions (excluding the nuisance parameter $f_\rho M$) were unimodal and skewed-Gaussian-like.}
To explore the effect of modeling choices on the outcome of the fits, we vary the following:

\vspace{0.3cm} 
\noindent \textit{Polytropic index $n$ for the model of \cite{Sapir2017}}: \hl{We generally do not expect the choice of polytropic index to substantially modify the results}. \autoref{sec:fitting:eq:sw17_t} and \autoref{eq:lt_sw17} indicate that the temperature and luminosity (respectively) are only weakly dependent on $n$. However, at times close to the transparency timescale $t_{\mathrm{tr}}$, emission interior to the edge of the envelope becomes visible and the emission becomes more sensitive to the density structure. Since IIb SNe are expected to have low envelope masses, $t_{\mathrm{tr}}$ is significantly reduced, increasing the expectation that the emission will be affected by choice of $n$.

\vspace{0.3cm} 
\noindent \textit{UV line suppression factor $A_{\mathrm{sup}}$ for the model of \cite{Morag2023}}: Equation A4 of \cite{Morag2023} presents a modification of the specific luminosity, introducing a factor $A_{\mathrm{sup}}=0.74$ to suppress UV line emission. This factor accounts for suppression caused by dominance of atomic transitions at shorter wavelengths. Given our dataset has multiple \textit{Swift} bands probing the UV emission of the transient, we may be sensitive to the suppression factor. We vary $A_{\mathrm{sup}}$ to be either $0.74$ or $1$ (no suppression) for the \cite{Morag2023} fits to test our sensitivity to the suppression factor. We additionally explore the effect of the UV data on the fits by performing a fit with $A_{\mathrm{sup}}=0.74$ but with the \textit{Swift} \textit{UVM1} and \textit{UVM2} filters removed from the dataset.  

\vspace{0.3cm} 
\noindent \textit{Presence of the earliest ATLAS detection}: The inclusion or exclusion of the earliest ATLAS detection at $t_{\mathrm{AT}}\sim59681.40$ has direct consequences for the inferred explosion epoch of the transient, which in turn affects what data can be included in the likelihood calculation according to the transparency timescale. For the purposes of investigating the impact of the rise-constraining ATLAS detection on the analysis, we have performed fits where the detection was omitted (``-ATLAS''), included (``+ATLAS'') with $t_{\mathrm{exp}}$ restricted from varying past $t_{\mathrm{AT}}$, and included with $t_{\mathrm{exp}}$ allowed to vary past $t_{\mathrm{AT}}$ (+ATLAS$^{a}$).

\vspace{0.3cm} 
\hl{We vary all the above considerations simultaneously, but only report a subset of combinations which is sufficient to describe the range of observed model behaviors. In particular, we focus on the model of \cite{Morag2023}, in order to assess the performance of this new model on Type IIb SNe. By contrast, the model of \cite{Sapir2017} serves as a reference point, and a way to compare to past IIb SNe measurements made using this model \citep[see, e.g.,][]{Arcavi2017,Pellegrino2023}. }
Finally, in addition to the cross-model comparison, we compare the most plausible model fit to best-fit parameters for other objects in the literature, with the intent of contextualizing SN 2022hnt to other Type II and IIb SNe. 
\section{Analysis and physical implications}
\label{sec:physics}

Having applied the framework of considerations discussed in \autoref{sec:fitting}, we now perform all proposed fits and assemble a landscape of parameter estimates.

\begin{table*}
    \begin{centering}
    \begin{tabular}{cccccccccc}
    \bottomrule
    Model& $n$ & $A_{\textrm{sup}}$ & ATLAS? & $v_{s,*}$ ($10^{8.5}$ cm/s) & $M_{\textrm{env}}$ ($M_\odot$) & $f_\rho$ M ($M_\odot$) & $R_0$ ($10^{13}$ cm) & $t_{\textrm{exp}}$ & $t_{\textrm{exp}} - t_{\textrm{AT}}$ (days) \\
    \hline
    SW17 & $3/2$ & N/A & Yes & $1.7^{+0.05}_{-0.04}$ & $0.10^{+0.04}_{-0.02}$ & $110\pm70$ & $0.64\pm0.09$ & $59681.1\pm0.05$ & $-0.3\pm0.06$ \\ \cdashline{1-10}
    SW17 & $3/2$ & N/A & No & $2.8^{+0.7}_{-0.6}$ & $0.11^{+0.04}_{-0.02}$ & $110\pm70$ & $0.20^{+0.12}_{-0.07}$ & $59682.4^{+0.3}_{-0.8}$ & $1^{+0.3}_{-0.8} $\\ \cdashline{1-10}
    SW17 & 3 & N/A & Yes & $1.7^{+0.13}_{-0.1}$ & $0.9^{+0.3}_{-0.2}$ & $90\pm80$ & $0.64\pm0.06$ & $59681.31\pm0.02$ & $-0.09\pm0.02$ \\ \cdashline{1-10}
    SW17 & 3 & N/A & No & $2.5^{+0.5}_{-0.7}$ & $1.1\pm0.4$ & $110\pm60$ & $0.19^{+0.34}_{-0.06}$ & $59682.5^{+0.2}_{-0.9}$ & $1.1^{+0.2}_{-0.9}$ \\ \cdashline{1-10}
    SW17 & $3/2$ & N/A & Yes$^{a}$ & $2.0^{+0.4}_{-0.2}$ & $0.12^{+0.06}_{-0.03}$ & $80\pm80$ & $0.55^{+0.09}_{-0.17}$ & $59681.27\pm0.05$ & $0.87\pm0.05$ \\ \cdashline{1-10} 
    M24 & $3/2$ & 0.74 & Yes & $3.6^{+0.2}_{-0.8}$ & $0.03^{+0.012}_{-0.008}$ & $50\pm40$ & $0.57\pm0.04$ & $59681.32\pm0.03$ & $-0.08\pm0.03$ \\ \cdashline{1-10}
    M24 & $3/2$ & 1 & Yes & $2.0^{+0.4}_{-0.3}$ & $0.022^{+0.013}_{-0.007}$ & $110\pm60$ & $0.6\pm0.1$ & $59681.32\pm0.03$ & $-0.08\pm0.03$ \\ \cdashline{1-10}
    M24 & $3/2$ & 1 & No  & $2.5\pm0.4$ & $0.05^{+0.02}_{-0.01}$ & $30\pm10$ & $0.5\pm0.2$ & $59681.7\pm0.05$ & $0.3\pm0.05$ \\ \cdashline{1-10}
    M24 & $3/2$ & 1 & Yes$^{a}$ & $2.3^{+0.4}_{-0.3}$ & $0.07\pm0.02$ & $24\pm14$ & $0.64^{+0.07}_{-0.09}$ & $59681.31\pm0.05$ & $-0.09\pm0.05$ \\ \cdashline{1-10}
    M24 & $3/2$ & 0.74 & No & $3.8^{+0.5}_{-0.5}$ & $0.018^{+0.006}_{-0.003}$ & $40\pm40$ & $0.22^{+0.07}_{-0.04}$ & $59682.8^{+0.1}_{-0.3}$ & $1.4^{+0.1}_{-0.3}$ \\ \cdashline{1-10}
    M24 & $3/2$ & 0.74 & Yes$^{a}$ & $3.1^{+0.5}_{-0.4}$ & $0.022^{+0.008}_{-0.005}$ & $110\pm6$ & $0.5^{+0.2}_{-0.1}$ & $59682.6\pm0.2$ & $1.2\pm0.2$ \\ \cdashline{1-10}
    M24 & $3/2$ & 0.74$^{b}$ & Yes & $2.7^{+0.4}_{-0.3}$ & $0.17^{+0.09}_{-0.06}$ & $3\pm2$ & $0.4\pm0.07$ & $59681.25^{+0.1}_{-0.2}$ & $-0.15^{+0.1}_{-0.2}$ \\ \cdashline{1-10}
    \toprule
    \end{tabular}
    \end{centering}
    $^{a}$ = the earliest ATLAS detection was included, but the explosion epoch was allowed to vary beyond it.\\
    $^{b}$ = fit performed with \textit{UVM1} and \textit{UVM2} filters removed.
    \caption{Numerical results from fitting the models described in \autoref{sec:fitting}.}
    \label{tab:results}
\end{table*}

\subsection{Cross-model analysis}

\begin{figure}
    \centering
    \includegraphics[scale=0.65]{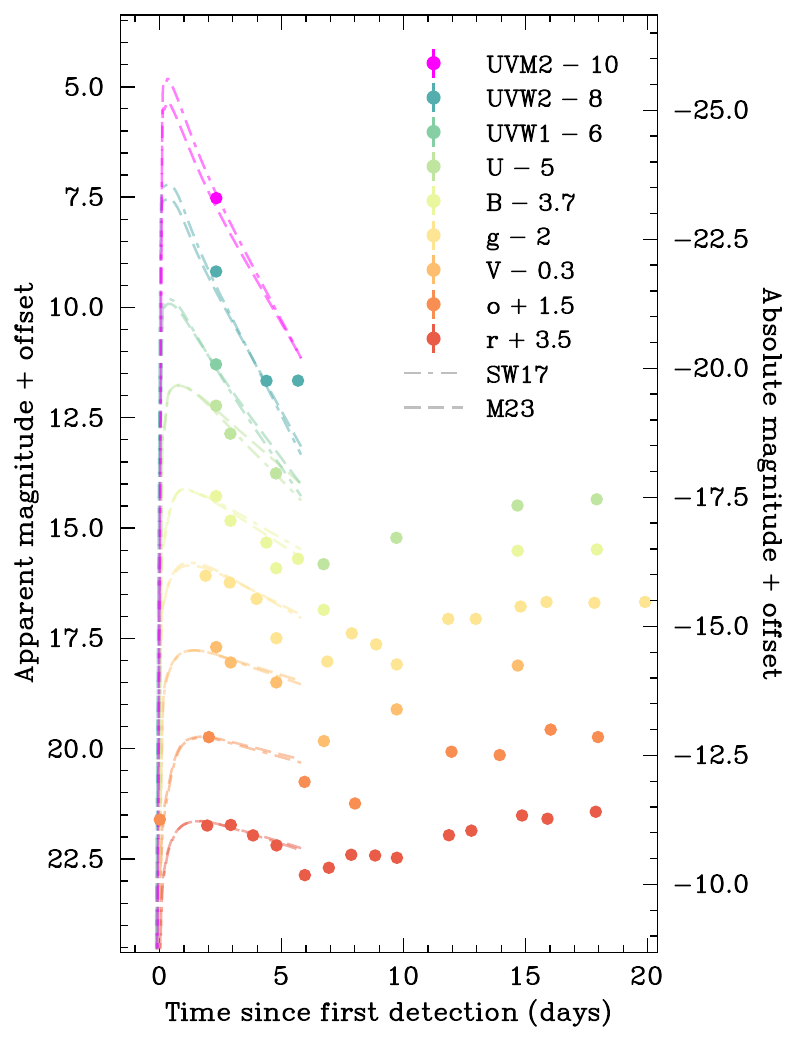}
    \caption{\hl{Comparison of representative maximum likelihood fits to the models of \cite{Sapir2017} (dot-dashed) and \cite{Morag2023} (dashed). The shock cooling models are extrapolated linearly beyond their regime of validity (up to day 6) to assess the performance beyond the transparency timescale. The shock cooling behavior is well-captured in all bands.}}
    \label{sec:implications:fig:early_comparison}
\end{figure}

The results of each fit are shown in \autoref{tab:results}. Representative fits for both the model of \cite{Sapir2017} and \cite{Morag2023} are shown in \autoref{sec:implications:fig:early_comparison}. 
Broadly, we find a wide range of envelope masses and explosion epochs, with a narrower range of shock velocities and progenitor radii. Envelope masses range from plausible IIb masses ($0.01 M_\odot \lesssim M_{\mathrm{env}} \lesssim 0.1 M_\odot$) to less plausible masses ($M\sim M_\odot$). These higher masses are more consistent with a Type II or short-plateau Type II explosion \citep[e.g.,][]{Lyman2016,Hiramatsu2021}, scenarios which are both inconsistent with the photometric and spectral evolution characteristics. Additionally, such a high mass is only estimated in models with polytropic index $n=3$, which has two ramifications. First, the high sensitivity of the mass estimate to $n$ indicates that we are in the low-mass ($M\lesssim 0.1 M_\odot$) regime, as discussed above. Second, given the other evidence that the transient is a IIb with a lower envelope mass, the higher mass estimate reported by the $n=3$ models indicates that the density profile is more consistent with a fully convective envelope ($n=3/2$). Indeed, the $n=3/2$ models report envelope mass estimates more plausible for a IIb SNe. We discuss only the $n=3/2$ models below, unless otherwise noted.

\hl{The model of \cite{Piro2021} extends the earlier model of \cite{Piro2015}, separating the envelope into two zones of variable power-law density to more precisely model the resulting shock cooling emission as the photosphere recedes (in mass coordinates) into the envelope. This model has been applied to several Type IIb SNe in the literature \citep[e.g.,][]{Arcavi2017,Yuhan2020,Pellegrino2023}. However, this model typically requires granular data within ${\sim}$several days of the shock cooling peak in order to produce meaningful fits. Our dataset has a significant gap of $\sim2.5$ days around the presumed shock cooling peak where no observations were made. As a result, there is likely at most one epoch within the regime of validity of the \cite{Piro2021} model. We attempted fits using this singular epoch (following the approach of \cite{Pellegrino2023}) and found the limited data to be insufficient to constrain any of the model features. We also unsuccessfully attempted fits of the \cite{Piro2021} model using data beyond the presumed regime of validity. As a result, we are choosing to omit the \cite{Piro2021} model from application to this particular object. }

\begin{figure*}
    \centering
    \includegraphics[scale=0.52]{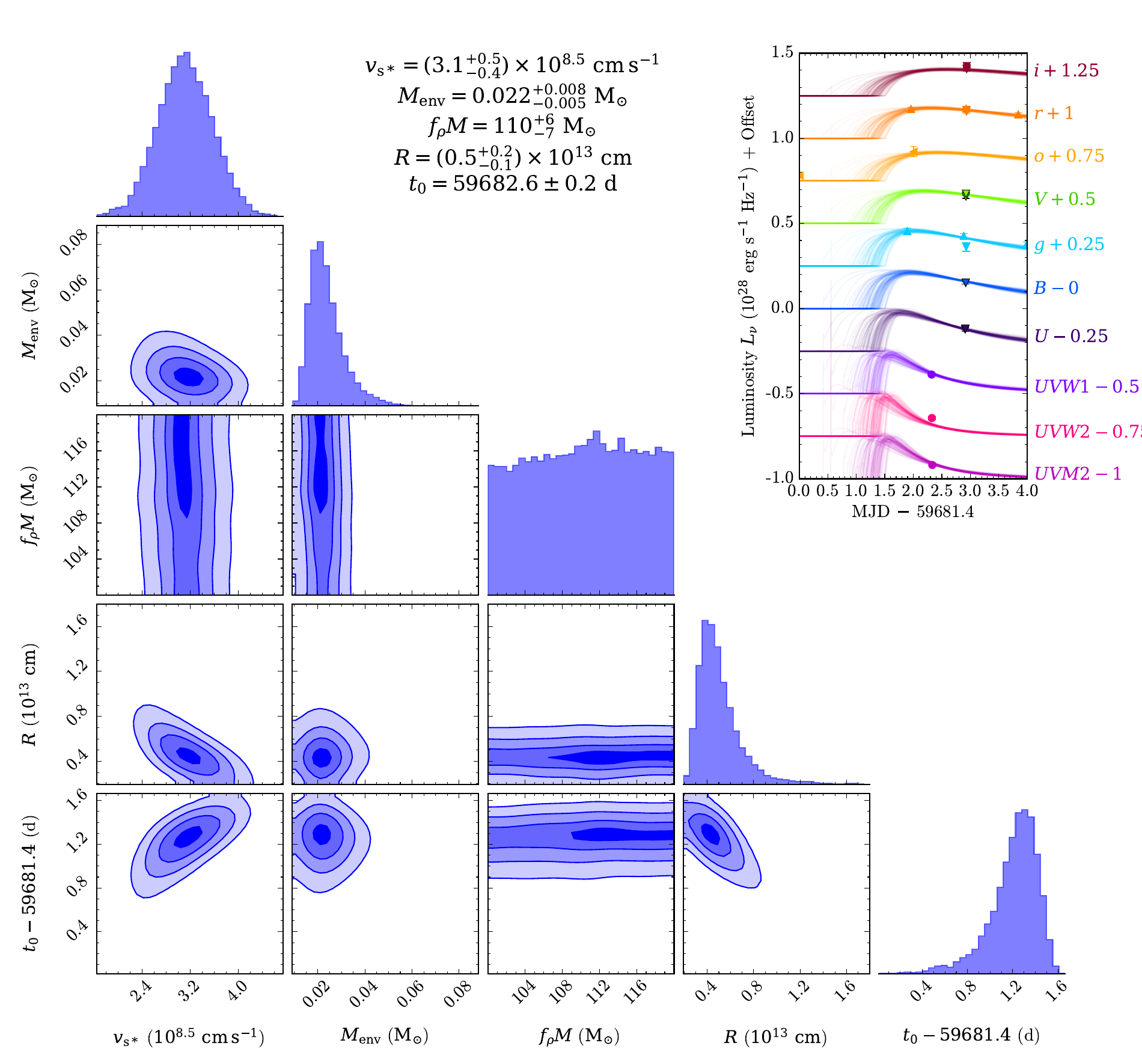}
    \caption{Corner plot and posterior samples for the \cite{Morag2023} model fit with $A_{\mathrm{sup}}=0.74$ and inclusion of the early ATLAS detection with allowed variation past $t_{\mathrm{AT}}$. The low envelope mass (and thus, short transparency timescale) results in the model clearly favoring an explosion epoch of $\sim1.25$ days after the ATLAS detection. This placement of the explosion epoch is representative of the \cite{Morag2023} model behavior when UV line suppression is included. All models seem to underestimate the \textit{UVW2} flux to the degree shown here. Note that $f_\rho$ is a nuisance parameter and is not expected to be well-constrained.}
    \label{sec:implications:fig:m24_uv07_ATLASsoft}
\end{figure*}
\begin{figure*}
    \centering
    \includegraphics[scale=0.56]{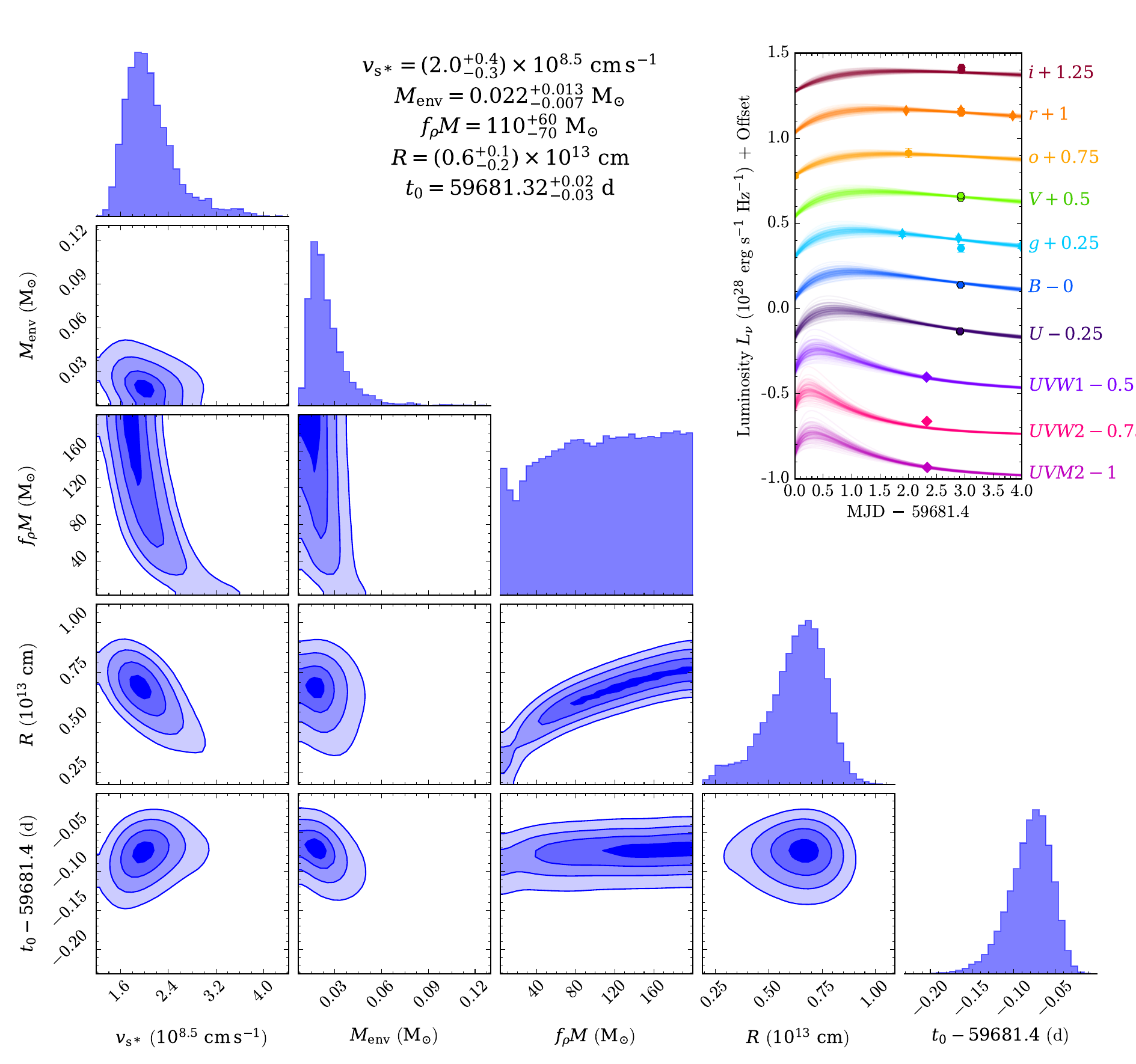}
    \caption{Corner plot and posterior samples for the \cite{Morag2023} model fit with $A_{\mathrm{sup}}=1.0$ and inclusion of the early ATLAS detection with variation past $t_{\mathrm{AT}}$ not allowed. The inclusion of the ATLAS detection clearly results in a strong constraint on the rise when the explosion epoch is not allowed to vary past $t_{\mathrm{AT}}$. All models seem to underestimate the UVW2 flux to the degree shown here. Note that $f_\rho$ is a nuisance parameter and is not expected to be well-constrained. }
    \label{sec:implications:fig:m24_uv1_ATLAShard}
\end{figure*}

Next, we examine shock velocity. \hl{The best-fit values from all models report a relatively narrow range of shock velocities, between $2.0\times10^{8.5} \mathrm{\ cm \ s^{-1}} \lesssim v_{s,*} \lesssim 4.0 \times10^{8.5} \mathrm{\ cm \ s^{-1}}$ ($6300 \mathrm{\ km \ s^{-1}} \lesssim v_{s,*} \lesssim 12700 \mathrm{\ km \ s^{-1}}$)}. \hl{The velocities estimated from the model fits are consistent to $1\sigma$ (model uncertainty) with the range identified from the He I spectra lines, which are not predictive of the initial shock velocity but are comparable.} Higher shock velocities ($v_{s,*} \gtrsim 3\times10^{8.5} \mathrm{\ cm \ s^{-1}}$) are exclusively generated in the \cite{Morag2023} model fits, which implement the UV line suppression with $A_{\mathrm{sup}}=0.74$. When the UV line suppression is not included (or included, but the shorter wavelength data is excluded), the shock velocity estimates drop to be more consistent with the model of \cite{Sapir2017}. Similarly, the inclusion of the UV line suppression factor in the \cite{Morag2023} model results in mass estimates $\sim10$x smaller than the \cite{Sapir2017} model or the \cite{Morag2023} model with no suppression. Visual examination of the \cite{Morag2023} fits both with (\autoref{sec:implications:fig:m24_uv07_ATLASsoft}) and without (\autoref{sec:implications:fig:m24_uv1_ATLAShard}) UV line suppression do not show clear differences in the ability to reproduce the UV photometry. Both fits are able to reproduce the \textit{UVW1} and \textit{UVM2} photometry, while simultaneously substantially underestimating the \textit{UVW2} photometry. However, the model including UV line suppression appears to slightly underestimate the $U$-band photometry compared to the model with no line suppression.

In addition to shock velocity, the effect of UV line suppression can be seen in the reported progenitor radius $R_0$. All fits report values for the progenitor radius in the range $30 R_\odot \lesssim R_0 \lesssim 100 R_\odot$, a relatively narrow range consistent with a yellow supergiant, which have been identified as the progenitors for some past IIb SNe \citep{2011dhYellow,2016gkgYellow,Niu2024}. \hl{Models with UV line suppression mostly favor a lower ($R \lesssim 60 R_\odot$) progenitor radius, with the sole exception being the model of \cite{Morag2023} with the early ATLAS detection included, which favors a progenitor radius of $\sim80 R_\odot$.} By contrast, all fits with the model of \cite{Sapir2017} favor a progenitor radius of $R_0\sim90 R_\odot$. \hl{In general, we found that mild posterior degeneracies existed between the progenitor radius and the other parameters. These did not improve with increased MCMC exploration time, and likely cannot be avoided. Indeed, we compared our posteriors to those reported in the literature \citep[e.g.,][]{Pellegrino2023,Hosseinzadeh2023} and identified similar degeneracies between the model parameters.}

Finally, we consider the best-fit explosion epoch. Most models clearly favor one of two explosion epochs: (i) $\sim1$ day after the ATLAS detection at $t_{\mathrm{AT}}$, or (ii) $\sim3$ hours before the ATLAS detection at $t_{\mathrm{AT}}$. There is little variation in these individual subgroups. The presence of the ATLAS detection at $t_{\mathrm{AT}}$ is an obvious but imperfect predictor of the estimated explosion epoch. \hl{When the ATLAS detection is present and the fit is forced to find an explosion epoch prior to it (+ATLAS), the fits prefer an explosion epoch $\sim3$ hours before $t_{\mathrm{AT}}$.} There is more variability for the other model configurations. If the ATLAS detection is excluded (-ATLAS), the explosion epoch is always preferred to be $\gtrsim0.25$ days after $t_{\mathrm{AT}}$. Results vary if the point is included and the fit is allowed to find an explosion epoch $>t_{\mathrm{AT}}$. We discuss the ATLAS detection at $t_{\mathrm{AT}}$ and its impact on the inferred explosion epoch in greater detail in \autoref{subsec:ATLAS}.

We now consider a fiducial set of best-fit parameters to use for comparison to other supernovae. Based on the above analysis involving the polytropic index, we immediately exclude all fits where $n=3$ from consideration. Given that the early ATLAS detection is only $\sim3\sigma$, we choose to exclude fits where the explosion epoch was forced to be $t_{\mathrm{exp}} < t_{\mathrm{AT}}$; i.e., we are restricting to the -ATLAS and +ATLAS$^{a}$ model configurations. Our next consideration is the UV suppression factor. Model configurations where $A_{\mathrm{sup}}=0.74$ universally favored mass estimates $M_{\mathrm{env}}\ll 0.1 M_\odot$, where the transparency timescale is $\lesssim1$ day. Such a short transparency timescale would exclude all but three data points, making a meaningful fit unlikely. By contrast, model configurations $A_{\mathrm{sup}}=1.0$ (i.e., no suppression) found masses consistent with a transparency timescale sufficient to achieve a meaningful fit; we therefore further restrict to these models. Of the remaining models (both \cite{Morag2023} and \cite{Sapir2017} where $A_{\mathrm{sup}}=1$ and ATLAS? is No or Yes$^{a}$), the parameters of interest ($v_{s,*}, M_{\mathrm{env}},R_0$) are largely consistent. \hl{As described in \autoref{sec:fitting:subsec:procedure}, we take the maximum a posteriori value of each posterior as our parameter estimate and calculate an uncertinaty using the highest posterior density interval. From this set of estimates $\mu_i$ and uncertainties $\sigma_i$ for each parameter, we calculate a fiducial weighted mean $p_{\text{fid}}$ and uncertainty $\sigma_{\text{fid}}$ using}
\begin{align}
    p_{\text{fid}} = \left[\sum_i \frac{1}{\sigma_i^2}\right]^{-1}\sum_i \frac{\mu_i}{\sigma_i^2},
\end{align}
and
\begin{align}
    \sigma_{\text{fid}} = \left[\sum_i \frac{1}{\sigma_i^2}\right]^{-1/2}.
\end{align}
Using this method, we find fiducial estimates for the parameters to be: $v_{s,*,f}=2.32\pm0.22\times10^{8.5}$ cm/s ($v_{s,*,f}=7.34\pm0.69\times10^{3}$ km/s), $M_{\mathrm{env},f}=0.068\pm0.013 \ M_\odot$, and $R_{0,f}=0.493\pm0.063\times10^{13}$ cm. 

\subsection{Comparison with other supernovae}

We additionally compare SN 2022hnt to samples from the spectrum of envelope-stripped SNe: specifically, SN 2016gkg (a typical stripped-envelope IIb), SN 2023ixf (a typical non-stripped-envelope Type II), and SN 2020bio (a peculiar IIb). These SNe were chosen in part due to their exceptionally early phase coverage. Both SN 2016gkg and SN 2020bio have data probing the shock cooling emission peak, and the dense SN 2023ixf follow-up campaign probed almost the entire shock cooling rise. Additionally, the SNe we have selected span from heavily-stripped-envelope objects (SN 2016gkg) to non-stripped-envelope objects (SN 2023ixf), in order to examine how the interpretation of the behavior of SN 2022hnt stands relative to similar objects.

\begin{figure}
    \centering
    \hspace*{-0.7cm}
    \includegraphics[scale=0.6]{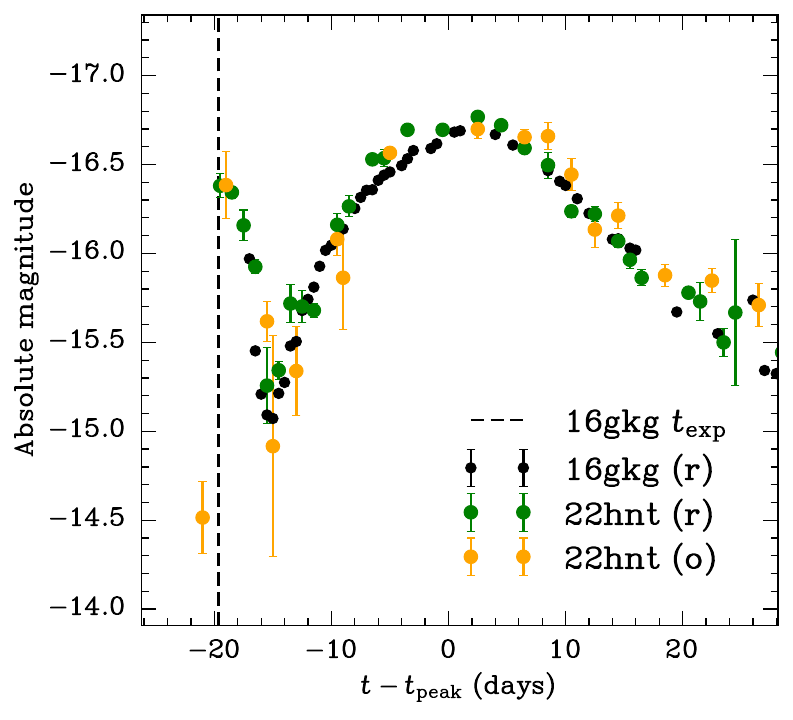}
    \caption{Comparison of the light curves of SN 2022hnt and SN 2016gkg. We compare the LCO data for 16gkg in $r$-band as well as the ATLAS data in $o$-band. The explosion epoch estimate (dashed line) for SN 2016gkg is based on \cite{Bersten16gkg}. Both light curves are shifted so the clearly-probed $^{56}$Ni peaks are aligned. SN 2022hnt appears to have a slightly slower rise to and fall from the shock cooling peak, indicating a higher envelope mass, consistent with the results from the shock cooling fits. } 
    \label{sec:imp:fig:16gkg}
\end{figure}

We begin by comparing the light curve of SN 2022hnt to SN 2016gkg, a similar Type IIb with a low envelope mass. Data were obtained from LCO archives and were published in \cite{Arcavi2017}. This visualization is made in \autoref{sec:imp:fig:16gkg}. To perform the comparison, we calculate absolute magnitudes for both objects and shift the light curves in time so that their Ni$^{56}$ peak in $r$-band is at $t=0$. The light curve peak is calculated via the maximum of a third-order polynomial fit to the nickel peak. We compare the photometric evolution in $r$-band as well as providing the SN 2022hnt $o$-band data to understand how the early ATLAS detection impacts the estimated explosion epoch. The light curves show remarkably similar evolution, particularly post-peak. Both light curves also show a prominent dip at $\sim14$ days pre-peak. However, SN 2016gkg drops significantly faster to this minimum from the shock cooling peak than SN 2022hnt does in either $r$- or $o$-band. This discrepancy could be attributed to the difference in the envelope masses for the two objects; SN 2016gkg was found to have an envelope mass of $\sim0.01 \ M_\odot$, while we find an envelope mass estimate for SN 2022hnt $\sim1$ order-of-magnitude larger. The larger envelope mass could plausibly result in a slower luminosity decay as more hydrogen is available for recombination. The early ATLAS detection is $\lesssim0.9$ hours away from the shifted best-fit explosion epoch for SN 2016gkg, providing a potential validation that the detection is related to the explosion and not an artifact due to random noise. Additionally, this comparison to another well-sampled IIb SNe emphasizes the uniquely early constraining power provided by such an early detection.

\begin{figure*}
    \centering
    \hspace*{-1cm}
    \includegraphics[scale=0.56]{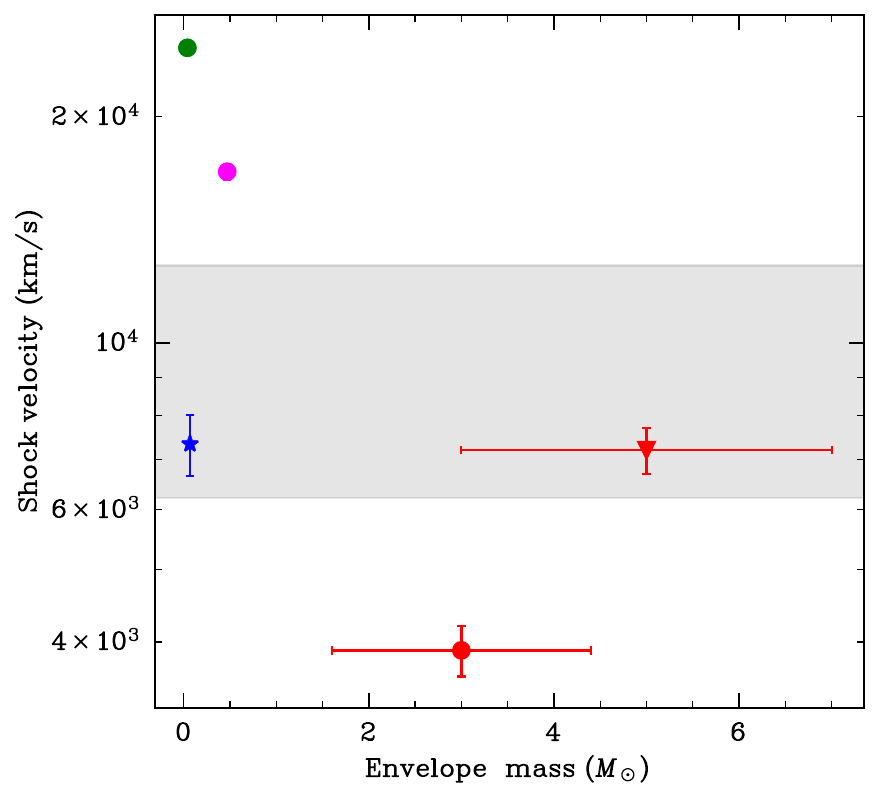}
    \includegraphics[scale=0.56]{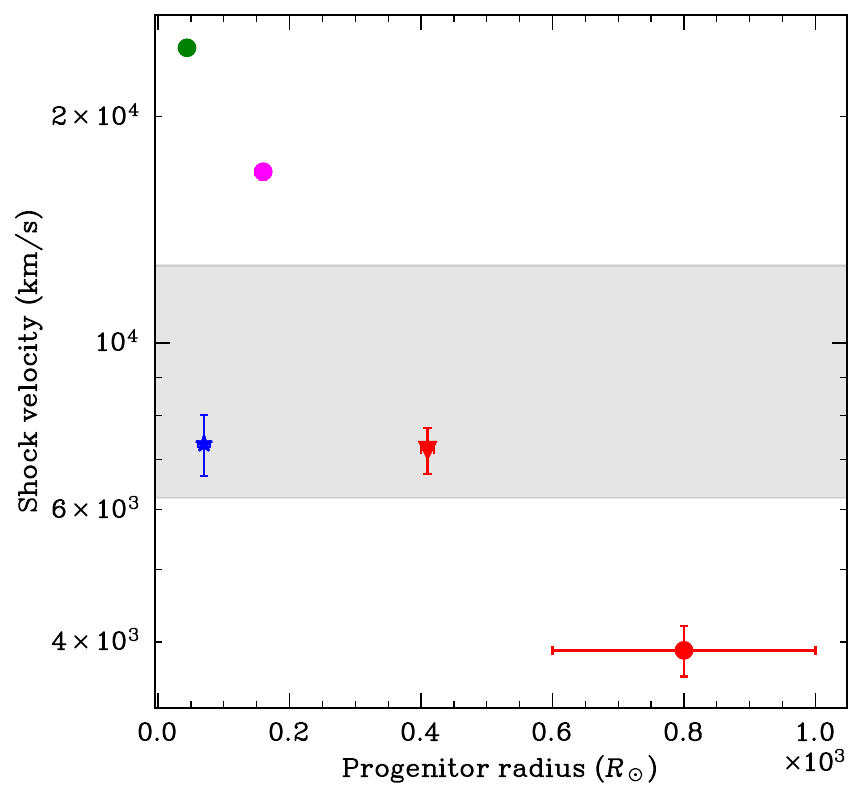}
    \hspace*{2cm}
    \includegraphics[scale=0.56]{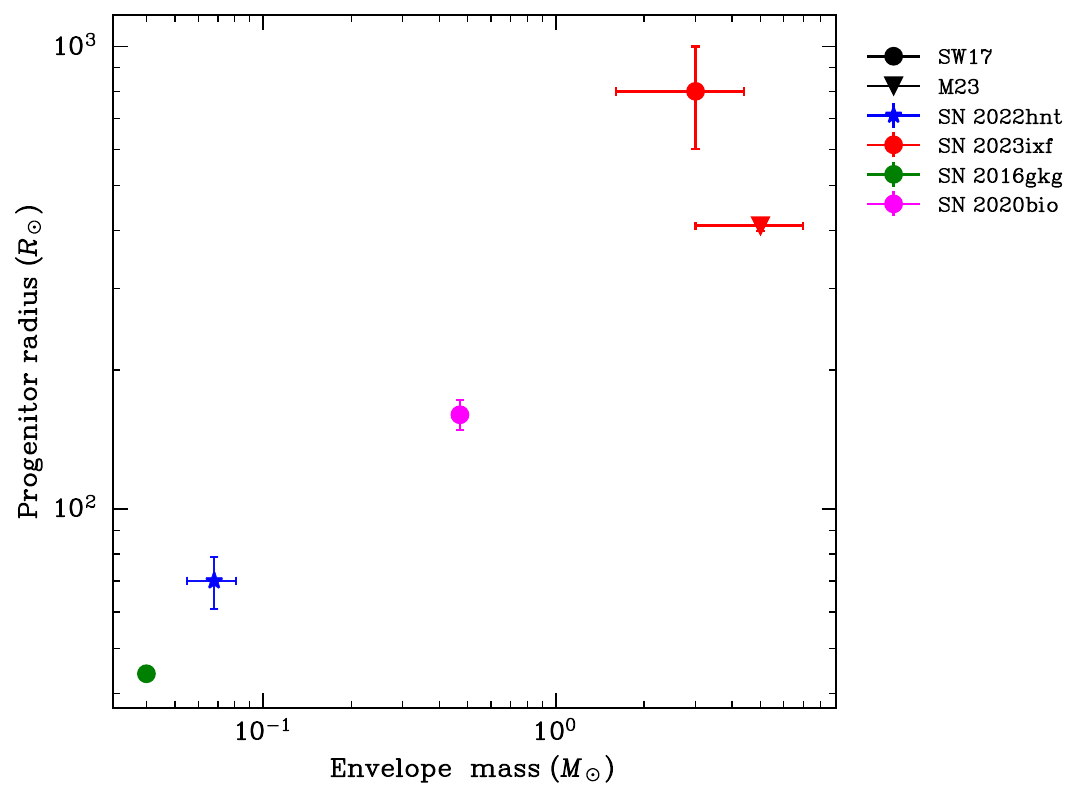}
    \caption{Cross-comparisons of shock velocity, envelope mass, progenitor radius, and rise time for SN 2022hnt, SN 2016gkg, SN 2023ixf, and SN 2020bio. Circles denote fits performed using the model of \cite{Sapir2017}, while triangles denote fits performed using the model of \cite{Morag2023}. The gray shaded region corresponds to the range of values recovered from the He I line velocity. Overall, we find shock velocities comparable with the He I line blueshift. The envelope mass reported by the shock cooling models is consistent with expectations from comparisons to other stripped and non-stripped envelope SNe (i.e., being less than Type II envelope masses but greater than very highly stripped envelope SNe such as 16gkg).}
    \label{sec:imp:fig:sn_comp}
\end{figure*}

Next, we compare the SN 2022hnt fiducial fit parameters to other objects in the literature. We visualize this comparison in \autoref{sec:imp:fig:sn_comp}. SN 2020bio and SN 2016gkg only have fits to the model of \cite{Sapir2017} available; however, we provide fits to both the model of \cite{Sapir2017} and \cite{Morag2023} for SN 2023ixf, as presented in \cite{Hoss2023}.

We first examine envelope mass, which is a direct probe of the pre-explosion stripping that differentiates Type IIb SNe from regular Type II SNe. The envelope of SN 2023ixf is considerably (i.e., $\gtrsim2$ orders of magnitude) more massive than either SN 2020bio, SN 2022hnt, or SN 2016gkg, appropriate for a Type II, which typically do not experience significant envelope stripping prior to explosion. SN 2022hnt and SN 2016gkg are more typical mostly-stripped-envelope Type IIb SNe, and as a result report envelope masses $\lesssim0.1 \ M_\odot$. SN 2020bio represents a more peculiar IIb; despite having extraordinarily weak $H\alpha$ and $H\beta$ emission features (indicating enhanced mass loss), shock cooling fits based on the model of \cite{Sapir2017} indicated a potentially less stripped envelope than more typical IIb SNe. However, early spectra indicated interactions with a CSM doped via significant mass loss, and models based on \cite{Piro2015} and \cite{Piro2021} reported envelope masses more consistent with a typical IIb.

We next examine progenitor radius. We generally find that the stripped-envelope objects in our comparison are consistent with our reported radius for SN 2022hnt, finding a progenitor radius of $\sim100 R_\odot$. This is substantially ($\sim1$ order of magnitude) less than radii reported from fits for SN 2023ixf, which is typical for Type II SNe.
We identify no clear relationship between progenitor radius and shock velocity. Aside from the model of \cite{Sapir2017} applied to SN 2023ixf--which reported a large but also quite uncertain progenitor radius--all progenitor radii in our comparison are relatively uniformly distributed below $\sim500 \ R_\odot$. Small ($R\lesssim200 \ R_\odot$) progenitor radii are identified in systems with fast (SN 2016gkg, SN 2020bio) and slow (SN 2022hnt, SN 2023ixf) estimated shock velocities.

We similarly do not identify a clear relationship between shock velocity and envelope mass. The shock velocity we obtain for SN 2022hnt is consistent with the He I line blueshift, and similar to the SN 2023ixf measured shock velocity as well as the He I line blueshift measured for SN 2020bio, which had a comparable progenitor radius and envelope mass. However, the shock cooling fits for both SN 2020bio and SN 2016gkg reported shock velocities a factor of $\sim2\textrm{-}3$ greater than SN 2022hnt. Unique among our fit parameters, the estimated shock velocity was largely statistically consistent regardless of model type or the inclusion of the earliest ATLAS detections. This consistency is in contrast to the model fits for SN 2023ixf and SN 2016gkg, which reported significant disagreement across the different classes of models used to measure shock velocity. However, despite reporting a shock velocity inconsistent with the He I line velocity, the modeled velocity measurements of SN 2020bio are self-consistent across model types. 

Finally, we consider explosion epoch. Assuming the detection at $t_{\mathrm{AT}}$ is non-spurious, we consider where such an early detection lies in the context of the literature. We consider three IIb SNe with early detections for comparison: (i) SN 2016gkg, (ii) SN 2017jgh, and (iii) SN 2020bio. For each of these three objects, we consider: (a) time of earliest detection relative to peak of the transient in $V$-band (denoted $\delta t_V$, inferred from shock cooling fits if needed), (b) time of the earliest detection relative to the estimated explosion epoch (denoted $\delta t_e$, inferred from shock cooling fits if needed), and (c) magnitude change between the earliest detection and the peak in the corresponding band (denoted $\delta m$). These values are summarized in \autoref{tab:early_detection_comp}. Immediately, we note a significant variation in the $\delta t_V$ estimate. Since all considered objects were discovered relatively shortly after explosion (i.e., $\delta t_e \ll \delta t_V$), the $\delta t_V$ quantity is mostly dominated by the rise time to the shock cooling peak, which is controlled by the envelope mass and shock velocity of the explosion. By contrast, the $\delta t_e$ quantity is more directly informative of how early the supernova was captured. The earliest ATLAS detection probes a regime similar to that probed by the earliest amateur detections of SN 2016gkg, with a comparable change in magnitude to peak ($\delta m$) as well. We demonstrate this with a visual comparison in \autoref{sec:imp:fig:16gkg}. 

\begin{table}
    \begin{centering}
    \begin{tabular}{cccc}
        \bottomrule
        Event & $\delta t_V$ (hours) & $\delta t_e$ (hours) & $\delta m$ \\ \hline
        SN 2016gkg$^{a}$ & -10.8 & 2.16 & 2.3 \\ \cdashline{1-4}
        SN 2017jgh & $-62$ & $<1^{b}$ & 4 \\ \cdashline{1-4}
        SN 2020bio$^{c}$ & -19.2 & 5.76 & 0.9 \\ \cdashline{1-4}
        \textbf{SN 2022hnt} & -28.8 & 2.88 & 2.1 \\ \cdashline{1-4}
        \toprule
    \end{tabular} \\
    \end{centering}
    $^{a}$We refer to \cite{Bersten16gkg} for these numbers.\\
    $^{b}$SN 2017jgh, remarkably, had Kepler data probing the entire rise from explosion as well as an extended sequence of non-detections immediately prior to the explosion epoch. We cite \cite{Armstrong2021} for these numbers.\\
    $^{c}$We refer to \cite{Pellegrino2023} for these numbers.
    \caption{Comparison of the earliest observations made in SN 2016gkg, SN 2017jgh, SN 2020bio, and SN 2022hnt.  \hl{$\delta t_V$ considers the time between the earliest observation and the shock cooling peak in $V$ band.} $\delta t_e$ considers the elapsed time between the explosion epoch and the earliest observation. $\delta m$ considers the magnitude change between the earliest detection and the peak in the same band.}
    \label{tab:early_detection_comp}
\end{table}

\subsection{Consequences of ATLAS constraint}
\label{subsec:ATLAS}



The earliest ATLAS detection at $t=t_{\mathrm{AT}}$ is a $\lesssim3$ sigma detection, which would conventionally be treated as ambiguous or even a non-detection. However, its proximity to the transient motivates an investigation into whether such emission would be plausibly predicted for a shock cooling event. The presence of the ATLAS detection has little impact on the best-fit envelope mass, progenitor radius, or shock velocity, an indicator that the emission at $t_{\mathrm{AT}}$ is consistent with the explosion parameters of SN 2022hnt. When the detection is removed, approximately 66\% of our model fits favor an explosion epoch $\gtrsim t_{\mathrm{AT}}$. However, the majority of the fits which favor $t_{\mathrm{exp}}\gtrsim t_{\mathrm{AT}}$ have $A_{\mathrm{sup}}=0.74$, which also produces a low envelope mass ($M_{\mathrm{env}}\sim0.01 \ M_\odot$) and therefore a very short transparency timescale ($t_{\mathrm{tr}}\sim0.5$ days). This unusually low mass estimate motivated the exclusion of the $A_{\mathrm{sup}}=0.74$ fits for consideration in the fiducial parameter set. Following that restriction, a significant majority of the remaining models prefer an explosion $\lesssim3$ hours prior to $t_{\mathrm{AT}}$. We therefore conclude that the emission at this epoch is consistent with shock cooling. Below, we consider the possibility that the emission is not related to the shock cooling (i.e., one of the models with $t_{\mathrm{exp}}\gg t_{\mathrm{AT}}$ are correct).

\section{Conclusions}
\label{sec:conclusions}

We report the follow-up campaign of SN 2022hnt, a Type IIb SNe at 84 Mpc. We organized an intense photometric and spectroscopic follow-up campaign shortly after discovery using the LCO telescope network. Additional data on SN 2022hnt were obtained using ZTF and ATLAS, including an exceptionally early detection in $o$-band by ATLAS. The presence of this early point constrains the shock cooling rise of the supernova and provides an opportunity to characterize the behavior of shock cooling models in a difficult-to-probe regime. 

Recently developed models of shock cooling can probe earlier in the supernova evolution than previous models, but require careful attention to a series of model considerations which affect performance and accuracy. We describe a framework for addressing the most significant considerations of existing shock cooling models into attempts to model the light curve, with the aim of producing conservative parameter estimates. This framework considers alternatives to shock cooling (e.g., heating due to wind breakout), intra- and inter-model consistency, and regimes of validity for individual models in order to construct a robust fiducial parameter estimate.

We used the models of \cite{Sapir2017} and \cite{Morag2023}, which characterize the emission generated by the rapid shock heating and cooling of the hydrogen envelope due to the supernova in the days following explosion. \hll{Both models have been used to provide measurements of envelope mass, shock velocity, progenitor radius, and explosion epoch for other objects; however, this marks the first application of the model of \cite{Morag2023} to a Type IIb SN.} We performed fits to the photometric evolution of SN 2022hnt using the \texttt{lightcurve-fitting} package, which employs an MCMC routine to produce maximum likelihood parameter estimates. To explore the impact of model choices on the parameter estimation, we varied polytropic index, degree of UV line suppression, and the presence of the earliest ATLAS detections. Despite the model of \cite{Morag2023} not being calibrated specifically for Type IIb SNe, we found that, regardless of model choice, it reports similar results to the more relevantly calibrated \cite{Sapir2017} model. We report fiducial estimates of the explosion properties of SN 2022hnt and find 
$v_{s,*,f}=7.08\pm0.44\times10^{3}$ km/s,
$M_{\mathrm{env},f}=0.07\pm0.02 \ M_\odot$, and $R_{0,f}=0.64\pm0.08\times10^{13}$ cm. Additionally, we characterize where these parameter estimates live on the spectrum of Type II supernovae. We find the shock velocity to be unusually low, but find a progenitor radius and envelope mass consistent with other Type IIb SNe and predictably lower than non-stripped-envelope Type II SNe.

The work presented here represents a framework that can be duplicated and expanded upon with more in-depth surveys of IIb SNe using the advanced shock cooling models that have been developed in recent years. \hll{Our treatment demonstrates that shock cooling models not specifically calibrated to IIb SNe \citep[such as][]{Morag2023} nevertheless can be cautiously extended to heavily stripped-envelope objects}. Future population studies of IIb SNe will more deeply explore the relationships between progenitor and explosion properties that have been presented here, and will better characterize the spectrum of IIb explosions. Our investigation of the impact of the early ATLAS detection demonstrates the importance of very early ($t\lesssim2$ days) data on Type IIb SNe, particularly in order to better constrain the effectiveness of the shock cooling assumption. The collection of early-time data as exploited here--and the rapid detection and follow-up that enables it--is crucial to test theoretical predictions and systematically study the behavior of IIb SNe. 

\vspace*{0.5cm}

\acknowledgements{JRF is supported by the National Science Foundation Graduate Research Fellowship Program under Grant No. (NSF 2139319). This work makes use of data from the Las Cumbres Observatory global telescope network.  The LCO group is supported by NSF grants AST-1911225 and AST-1911151. This research was supported in part by grant NSF PHY-2309135 to the Kavli Institute for Theoretical Physics (KITP). LJP is supported by a grant from the Simons Foundation (216179, LB).
}

\bibliography{ref}


\end{document}